\documentclass[12pt,a4paper]{article}
\usepackage[utf8]{inputenc}
\usepackage[T1]{fontenc}
\usepackage{amsmath,bbm,cite}
\usepackage{amsfonts}
\usepackage{amssymb,comment}
\usepackage{graphicx}
\usepackage{transparent}
\usepackage{notoccite}
\usepackage{ulem}
\usepackage[left=2cm, right=2cm]{geometry}
\usepackage{todonotes}
\usepackage{xcolor}
\usepackage{physics}
\usepackage{bbold}
\usepackage{ntheorem}
\usepackage{setspace} 
\usepackage[overlay,absolute]{textpos}
\usepackage{hyperref}

\newcommand{\be}{\begin{equation}}
	\newcommand{\ee}{\end{equation}}
\newcommand{\bea}{\begin{eqnarray}}
	\newcommand{\eea}{\end{eqnarray}}

\newcommand{\vol}{\mathcal{V}}
\newcommand{\HSpace}{\mathcal{H}}

\newcommand{\Z}{\mathbb{Z}}

\definecolor{AW}{HTML}{629202}
\definecolor{AH}{HTML}{d91f05}
\definecolor{SZ}{HTML}{0394f9}
\definecolor{BF}{HTML}{f903d7}

\newcommand\PlaceText[3]{%
\begin{textblock*}{10in}(#1,#2)
#3
\end{textblock*}
}%

\begin{document}
\sloppy

\PlaceText{160mm}{15mm}{DESY 24-175}
\PlaceText{160mm}{20mm}{18 Nov 2024}
\begin{center}
{\LARGE\bf End-of-the-World Branes and}

\vspace{.5cm}
{\LARGE\bf  Inflationary Predictions}

\vspace{.5cm}
{\LARGE\bf for Rocky and Swampy Landscapes}
\vspace{.5cm}
\end{center} 

\vspace*{0.8cm}
\thispagestyle{empty}

\centerline{\large 
Bjoern Friedrich$^{1}$,
Arthur Hebecker$^1$, Alexander Westphal$^{2}$}
\vspace{0.5cm}
 
\begin{center}
$^1${\it Institute for Theoretical Physics, Heidelberg University,}
\\
{\it Philosophenweg 19, 69120 Heidelberg, Germany}\\[0.5ex]  
\end{center} 
\vspace{-.4cm}
 
\begin{center}
$^2${\it Deutsches Elektronen-Synchrotron DESY,}
\\
{\it Notkestr. 85, 22607 Hamburg, Germany}\\[0.5ex]  
\end{center} 
\vspace{.25cm}
\centerline{\small\textit{E-Mail:} \href{mailto:friedrich@thphys.uni-heidelberg.de}{friedrich@thphys.uni-heidelberg.de}, \href{mailto:a.hebecker@thphys.uni-heidelberg.de}{a.hebecker@thphys.uni-heidelberg.de},}
\centerline{\href{mailto:alexander.westphal@desy.de}{alexander.westphal@desy.de}}

\vspace*{1.2cm}
\begin{abstract}\normalsize
Making cosmological predictions in a multiverse is a fundamental theoretical challenge. Assuming that (quasi-)de Sitter vacua are quantum mechanically described by a finite-dimensional Hilbert space, we develop a detailed framework for making explicit anthropic predictions. A key challenge which we attempt to overcome arises because, almost unavoidably, cosmologies that asymptote to 
Minkowski space exist.
We then apply our framework to predicting the scale of inflation.
We find that, even if eternal inflation is allowed,
our 
predictions depend on creation rates of universes from nothing. These, in turn, are highly sensitive to the existence of end-of-the-world branes.
The rates for the creation of universes from nothing are the dominant ingredient for `Swampy Landscapes', which may have no
metastable de Sitter vacua but only slow-roll solutions.
In `Rocky Landscapes', where long-lived de Sitter vacua are abundant, tunneling rates between such vacua represent a further key factor for deriving predictions. 

\vspace*{.4cm}
\noindent
\end{abstract}
\vspace{10pt}

\newpage
 
	\sloppy
\title{End-of-the-World Branes, Predictions in the String Landscape and the Cosmological Central Dogma}
\date{}
\author{}
\begin{spacing}{0.85}
\setcounter{tocdepth}{2}
    \tableofcontents
\end{spacing}

\section{Introduction}
For theories exhibiting different vacua, a measure on their set is required to achieve predictivity.
If these vacua are continuously populated by eternal inflation, by creation from nothing, or by some other process, this leads to a formidable theoretical challenge known as the measure problem.
This was initially studied in the context of eternal inflation \cite{Linde:1993nz} and became increasingly more prominent with the advent of the string landscape \cite{Bousso:2000xa, Kachru:2003aw, Susskind:2003kw, Denef:2004ze, Balasubramanian:2005zx} (see \cite{Schellekens:2015zua, Hebecker:2020aqr} for reviews). The reason is the initial expectation that de Sitter (dS) vacua are abundant and  
eternal inflation \cite{Steinhardt:1982kg, Vilenkin:1983xq, Guth:2007ng} is natural.
This perspective has changed more recently with the realization that stringy dS solutions are very difficult to realize and the proposal that they may be entirely absent \cite{Danielsson:2018ztv, Obied:2018sgi, Garg:2018reu, Ooguri:2018wrx, Dvali:2018jhn}.
Earlier studies have argued that dS is problematic for exponentially long lifetimes, see e.g.  \cite{Ford:1984hs,Mottola:1984ar,Antoniadis:1985pj,Tsamis:1992sx,Goheer:2002vf, Polyakov:2007mm,Dvali:2013eja, Dvali:2014gua, Dvali:2017eba}.
On the other hand, long-lived dS is also taken seriously as a quantum system and studied intensely  \cite{Banks:2000fe,Witten:2001kn,Strominger:2001pn,Spradlin:2001pw,Chandrasekaran:2022cip,Ivo:2024ill}.
In our present work, we will be agnostic concerning the abundance and longevity of dS vacua. In other words, we will allow both for `swampy’ landscapes  (using
the terminology of \cite{Blanco-Pillado:2019tdf}) in which dS is either extremely rare or non-existent, and `rocky’ landscapes with abundant dS vacua.

One approach to the measure problem, developed for eternally inflating cosmologies, is
to count measurements
before some late-time
cutoff which
regularizes the infinities, see e.g.~ \cite{Linde:1993nz, Linde:1993xx, Garriga:2005av, Bousso:2006ev, Susskind:2007pv, DeSimone:2008bq, Bousso:2008hz, Garriga:2008ks, Harlow:2011az}. 
However, the 
cutoff dependence makes this  somewhat arbitrary, as explained in the reviews \cite{Freivogel:2011eg, Vilenkin:2006xv}.
Other approaches appeal to quantum cosmology \cite{Vilenkin:1994ua, nomura2011physical,Bousso:2011up , Hertog:2011ky, nomura2012static,Hartle:2016tpo} or count measurements along the worldline of a single observer moving through the multiverse \cite{Dyson:2002pf, garriga2013watchers, nomura2011physical, nomura2012static}.
More recent work on the measure problem includes \cite{Carifio:2017nyb, Chiang:2018dju, Finn:2018krt, Jain:2019gsq, Khoury:2019yoo, Vilenkin:2019mwc, Khoury:2019ajl,Blanco-Pillado:2019tdf, Kartvelishvili:2020thd,Olum:2021pux,Khoury:2021grg,Khoury:2022ish,Khoury:2023ktz,Friedrich:2022tqk,Ivo:2024ill}.

In what follows, we will focus on a measure for the multiverse proposed in \cite{Friedrich:2022tqk}, which rests on the `cosmological central dogma' (CCD) \cite{Talks,Banks:2000fe,Susskind:2021yvs}: The CCD states that de Sitter (dS) space represents a finite quantum mechanical system with Hilbert space dimension $\exp(S_{dS})$, where $S_{dS}$ is the dS entropy. In the following, we assume that the CCD applies to all solutions with a cosmological horizon (quasi-dS solutions), including inflation and quintessence. For simplicity, we will refer to all solutions of this kind simply as `dS vacua'.
Conceptually, the proposal of \cite{Friedrich:2022tqk} is inspired by the quantum mechanical approach of \cite{Bousso:2011up,nomura2011physical,nomura2012static}, formalizing the latter and trying to develop it to a point where explicit predictions can be derived. The total Hilbert space is taken to be a direct sum, $\mathcal{H}=\bigoplus_{i\in vac} \mathcal{H}_i$, where $i$ labels the vacua of the underlying landscape.
This sum contains both finite-dimensional dS as well as infinite-dimensional Minkowski and AdS subspaces.  Furthermore, it is assumed that there exists a unique quantum state of the multiverse, $\Psi\in \mathcal{H}$, obtained by solving the Wheeler-DeWitt (WDW) equation with a source term. This source term characterizes the creation of universes from nothing as well as bubble-of-nothing decays. One may think of the proposal \cite{Friedrich:2022tqk} as a `local WDW measure', where the term `local' refers to the fact that only a local, horizon-sized patch of each dS vacuum is relevant in the multiverse dynamics.  Predictions can be made by constructing projection operators $P_\alpha,P_\beta$ on subspaces of interest and evaluating ratios
\begin{align}
    \frac{\bra{\Psi}P_\alpha\ket{\Psi}}{\bra{\Psi}P_\beta\ket{\Psi}}\,.\label{Prediction_Intro}
\end{align}
Note that only relative probabilities of the form \eqref{Prediction_Intro} can be calculable and physically meaningful. However, not all such ratios are a priori well-defined since, due to the presence of infinite-dimensional ${\cal H}_i$, the wave function $\Psi$ is in general not normalizable.

Clearly, the structure of the underlying Hilbert space and the associated observables depend on UV physics, such that a direct construction of $\Psi$ and the relevant projection operators appears impossible. 
However, the structure of $\Psi$ can partially be determined semiclassically \cite{Friedrich:2022tqk}:  The probability distribution of dS vacua in the landscape may be obtained by solving an appropriate rate equation.  The required input parameters are various vacuum creation, decay and transition rates.  This method does not allow to ascribe probabilities to Minkowski/AdS vacua since the projection of $\Psi$ on such vacua is in general not normalizable.

We note that similar rate equations have appeared before in the context of `local measures', in particular the `causal patch measure' \cite{Bousso:2006ev}. Such measures are defined by counting events along the worldline of an observer
\cite{Garriga:2005av, Bousso:2006ev, Bousso:2008hz, Bousso:2010im, garriga2013watchers}. The rate equations from 
\cite{Garriga:2005av,Bousso:2006ev, garriga2013watchers} are closest to our results. Nevertheless, the local WDW measure differs from this class since it is formulated observer-independently, on the basis of quantum mechanical principles. Observers make their appearance only once projections on appropriate Hilbert subspaces are introduced to derive anthropic predictions. Some of the differences between traditional local measures and the local WDW measure will play a key role in this work. The measure proposal of \cite{nomura2011physical, nomura2012static}, while conceptually related to our approach, also depends on an a priori chosen observer. Additionally, it does not involve the creation of vacua from nothing, which is by contrast of central importance in our analysis.

In this work, we explore the phenomenology and fundamental issues of the local WDW measure. It is clear that
the finiteness of predictions of the form \eqref{Prediction_Intro} is under threat because $\Psi$ is in general not normalizable. 
However, when considering anthropic predictions, the projection operators in \eqref{Prediction_Intro} need to involve a projection on observers, which may restore finiteness.
Since, arguably, the leading paradigm for creating anthropic observers is a period of inflation followed by reheating and structure formation, 
we focus on such corners of the landscape.

If observers live in asymptotic dS vacua, i.e.~if a (small) cosmological constant is present, we can approximate the observer projection as follows: First, we project on the pure dS-part of the Hilbert space which is finite dimensional and hence leads to well-defined predictions of the form \eqref{Prediction_Intro}.
For observers formed after a reheating period, we then project further on the subspace of states describing inflation.
We conclude that the probability $p({\rm obs},i)$ of an observer living in vacuum $i$ is, to leading order, given by
\begin{align}
    p({\rm obs},i)\propto p_{\rm inf (i)}\,,\label{Intro_p_o_i}
\end{align}
the probability of finding the universe in an inflationary state leading to reheating in vacuum~$i$. This latter probability may be obtained by treating inflationary plateaus as additional, short-lived dS vacua and then deriving their probability from a rate equation.

The situation changes if we consider landscapes with anthropically viable terminal vacua, i.e.~AdS or Minkowski. 
While our method to obtain relative probabilities for the relevant inflationary plateaus still works, a reheating process ending in Minkowski space may now lead to infinitely 
many observers in a causally connected region.\footnote{Since cosmological AdS always crunches, the analogous infinite set of observers is not contained in a causally connected patch.}
It hence appears that anthropic predictions could become ill-defined, due to both numerator and denominator of \eqref{Prediction_Intro} being infinite. 
A similar problem arises for the causal patch measure \cite{Bousso:2006ge, Freivogel:2011eg}.
A main result of our work is the analysis of several approaches for obtaining well-defined predictions even if observers in Minkowski vacua exist.

One of our suggested solutions is based on the observation that, on the reheating surface, it is not yet visible whether late-time evolution will lead to dS or Minkowski space. Hence, as in the CCD, not all Minkowski observers can be considered independent. 
In other words, the CCD suggests that there exist gauge redundancies on the reheating surface, such that only part of it can be seen as fundamental.
Then, the probability for an observer living in Minkowski space is again estimated by \eqref{Intro_p_o_i}, with ${{\rm inf}(i)}$ now denoting an inflationary plateau reheating into a Minkowski vacuum $i$. Another possibility is to embrace the infinity of observers on a Minkowski-space reheating surface as meaningful. This leads to the prediction that our observed dark energy must be of quintessence type.

Given the vacuum creation and transition rates, we can use the rate equation to derive the relative probabilities for finding the universe on different inflationary plateaus. The required transition rates can be calculated using the Coleman-de Luccia method \cite{coleman1980gravitational}. We do so specifically for KKLT- and LVS-type vacua \cite{Kachru:2003aw, Balasubramanian:2005zx}, which we thus use as an example of a possible `rocky' landscape.
Furthermore, vacua can decay to nothing. This was first demonstrated by Witten \cite{Witten:1981gj} and has then been further studied in e.g. \cite{Blanco-Pillado:2011fcm,Blanco-Pillado:2016xvf,GarciaEtxebarria:2020xsr,Draper:2021qtc,Bomans:2021ara,Angius:2022aeq}.
Recently \cite{Friedrich:2023tid}, decays to nothing in the string landscape were analyzed explicitly.

Our approach to the measure problem is only consistent if, in addition to tunneling transitions, finite-size universes can be created from nothing via a quantum process. Proposals exist for the creation of spherical universes \cite{Vilenkin:1982de, Vilenkin:1983xq, hartle1983wave, Linde:1983mx, Vilenkin:1984wp} (cf.~also the new proposal of \cite{Friedrich:2024aad}), universes of disk topology with a boundary \cite{Hawking:1998bn,Turok:1998he,Garriga:1998ri,Bousso:1998pk,Blanco-Pillado:2011fcm,Friedrich:2023tid} and closed universes of nontrivial topology \cite{Zeldovich:1984vk,Coule:1999wg,Linde:2004nz}.
There remains an ambiguity concerning the sign with which the instanton action appears in the exponent of the corresponding creation rates (the Hartle-Hawking vs. Linde/Vilenkin sign choice). This is related to different possible definitions of the gravitational path integral, cf. \cite{Lehners:2023yrj} and refs.~therein. 
In a recent paper \cite{Ivo:2024ill}, it was suggested to derive the local density matrix using the Hartle-Hawking proposal. It would be interesting to find a connection of this to the local WDW measure.
Phenomenologically, the Hartle-Hawking choice appears to be strongly disfavored as it predicts the shortest anthropically viable period of inflation, in conflict with observation (cf.~the recent discussion in~\cite{Maldacena:2024uhs}).

Since we assume that observers form only near reheating surfaces, the scale of the preceding inflationary period is an important observable. It becomes a natural target for an anthropic prediction because it does not directly impact our existence. This prediction also assumes that, within the underlying landscape, the inflationary scale is largely uncorrelated with anthropically significant parameters, such as certain Standard Model couplings.
In the local WDW framework, the prediction turns out to strongly depend on vacuum creation processes. This is particularly apparent for `swampy' landscapes. In this case, tunneling between vacua is subdominant and the probability of finding ourselves in a given vacuum is mainly determined by its probability to nucleate from nothing.
In determining the vacuum creation rates, the sign choice when relating the instanton action to the decay rate, the topology of created universes, and the tensions of ETW branes existing in the landscape play a significant role.
Nevertheless, it is possible to establish close links between the gravitational path integral and cosmological predictions. While the Hartle-Hawking sign choice already seems to be ruled out, we argue that the Linde/Vilenkin proposal points to primordial gravitational waves that are detectable by upcoming CMB-experiments.

This article is organized as follows: In Sect.~\ref{sect_Local_WDW_review}, we briefly review the local WDW measure and explain how elementary anthropic predictions can be made.
In Sect.~\ref{sect_projecting_on_observers}, we discuss the problem of infinitely many observers living in terminal vacua and provide possible resolutions.
In Sect.~\ref{sect_strat}, we discuss the fundamental ingredients needed to predict the most likely scale of inflation to be measured by anthropic observers.
In Sect.~\ref{sect_String_rates}, we explicitly calculate various vacuum decay rates for KKLT and LVS type vacua. 
In Sect.~\ref{sect_explicit_predictions}, we combine the results of the previous sections to try and make an explicit prediction for the scale of inflation relevant for anthropic observers in a string multiverse.

\section{The local WDW measure and anthropic predictions}\label{sect_Local_WDW_review}
\subsection{Review of the local WDW measure}
\begin{figure}[ht]
	\centering
    \def\svgwidth{0.5\linewidth}
	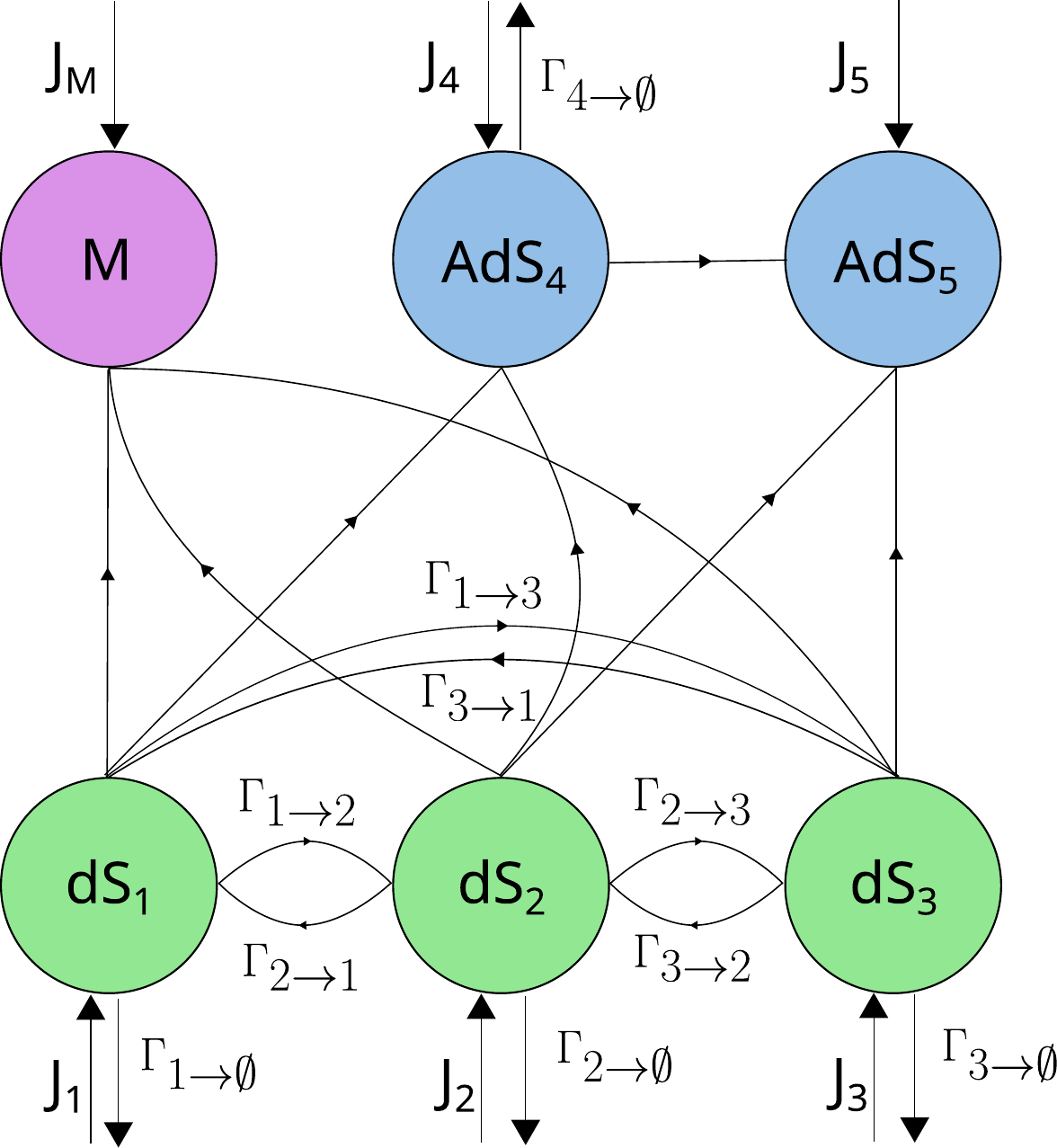
	\caption{Creation and tunneling for a toy landscape consisting of three dS, one supersymmetric Minkowski (M), one non-supersymmetric and one supersymmetric AdS vacuum (`AdS$_4$' and `AdS$_5$' respectively). Note that the labels distinguish different vacua and have nothing to do with dimensionality, which is always 4. To avoid overburdening notation, we have not put labels on the arrows for decays to Minkowski and AdS vacua.}
	\label{fig_tunneling}
\end{figure}
It is our goal to implement a measure for the multiverse based on the following assumptions:
\begin{itemize}
    \item The `cosmological central dogma' (CCD) \cite{Banks:2000fe,Susskind:2021yvs} holds. In other words, de Sitter (dS) space has to be viewed from a static patch perspective and its Hilbert space $\HSpace_{dS}$ has dimension $\dim(\HSpace_{dS})=\exp(S_{dS})$, where $S_{dS}$ is the dS entropy. The CCD may be interpreted as the statement that different static patches of dS space are gauge-redundant.
    \item The state vector $\Psi$ or wave function of the multiverse lives in the Hilbert space
    \begin{align}
        \mathcal{H}=\bigoplus_{i\in vac}\mathcal{H}_i\,,\label{Hilbert_space}
    \end{align}
    with each landscape vacuum $i$ contributing a subspace $\mathcal{H}_i$. These subspaces are finite-dimensional for dS and infinite-dimensional for AdS and Minkowski vacua.
    \item The wave function of the universe $\Psi$ solves a modified Wheeler- (WDW) equation 
    \begin{align}
    H\Psi=\chi \label{WDW_source}\,.
    \end{align}
    Here the source term $\chi$ characterizes the creation of universes from nothing as well as the vacuum decay to nothing.
\end{itemize}

Solving \eqref{WDW_source} is certainly very difficult since the microscopic structure of the Hamiltonian is unknown.  However, as argued in \cite{Friedrich:2022tqk}, our assumptions stated above imply that the probability distribution corresponding to $\Psi$ can be derived semiclassically for the dS part of the multiverse: Specifically, we denote by 
\begin{align}
    p_i=\norm{P_i\Psi}^2\label{p_i_Psi_def}
\end{align}
the probability for finding a vacuum of type $i$ in the multiverse. 
Here, $P_i$ is the operator projecting on the subspace $\mathcal{H}_i$. The probabilities $p_i$ obey the semiclassical rate equation
\begin{align}
		J_i=\sum_{j\,\in\,dS} \left(p_i\Gamma_{i\to j}-p_j\Gamma_{j\to i}\right)\,+\,\,\,p_i\!\!\!\sum_{y\,\in\, Terminals}\Gamma_{i\to y}\, ,\label{rate_equation}
	\end{align}
where $i,j$ run over all dS vacua, and $\Gamma_{i\to j},\Gamma_{i\to y}$ and $J_i$ represent the vacuum transition and creation rates, respectively. The index $y$ runs over terminal vacua, i.e.~AdS or Minkowski, and `nothing'.
Similar rate equations have appeared in \cite{Garriga:2005av,garriga2013watchers}.
The wave function $\Psi$ can be interpreted as characterizing a stationary state with universes constantly being created, decaying and tunneling into each other according to Eq.~\eqref{rate_equation}. We have attempted to illustrate this multiverse dynamics in Fig.~\ref{fig_tunneling}.

With the normalization of $\Psi$ unknown, only ratios of probabilities are meaningful. As already shown in Eq.~\eqref{Prediction_Intro}, they take the form
\begin{align}
    \frac{\bra{\Psi}P_{\alpha}\ket{\Psi}}{\bra{\Psi}P_{\beta}\ket{\Psi}}\,,\label{Local_WDW_prediction}
\end{align}
where $P_\alpha,P_\beta$ project $\Psi$ on subspaces of interest, e.g.~two different vacua. The ratio of their expectation values specifies the relative probability for condition `$\alpha$' being realized rather than `$\beta$'.

Since a direct construction of $\mathcal{H}$ and the projection operators $P_\alpha,P_\beta$ seems unfeasible, we will focus on predictions which can be obtained from semiclassical considerations by using Eq.~\eqref{rate_equation}.
While vacuum transition rates $\Gamma_{i\to j}$ can be calculated using the Coleman-de Luccia method \cite{coleman1980gravitational}, there is a larger uncertainty regarding the creation rates $J_i$.\footnote
{We shall comment on all these computations in Sect.~\ref{sect_strat}.}
Most commonly employed are the Hartle-Hawking no-boundary \cite{Hawking:1998bn} and the Linde/Vilenkin tunneling \cite{Linde:1983mx, Vilenkin:1984wp} proposals, the predictions of which are already drastically different, due to a different sign appearing in the exponent when relating the instanton action to the creation rate.
Additionally, a creation via a `bubble of something' has been proposed \cite{Hawking:1998bn,Turok:1998he}, and a non-trivial topology in the nucleated universe can play an important role \cite{Zeldovich:1984vk,Coule:1999wg,Linde:2004nz}.
Finally, it was speculated that also the creation of universes with a boundary in the off-shell region may be relevant \cite{Friedrich:2024aad}.

\subsection{Anthropic predictions from the local WDW measure for observers in dS universes}
\label{sect_anthropic_pred_WDW}
We now address the `anthropic questions' of the most likely vacuum for us to live in and of the inflationary scale we are most likely to observe, following the strategy of \cite{Friedrich:2022tqk}. As discussed in the introduction, we shall focus on anthropic observers originating on reheating surfaces after a period of inflation.
The probability $p({\rm obs},i)$ for such an observer to be found in vacuum $i$ is then certainly proportional to $p_{\rm inf (i)}$, the probability of finding the universe in an inflationary state `inf$(i)$' leading to reheating in vacuum $i$. We hence write (cf.~Eq.~\eqref{Intro_p_o_i})
\be
p({\rm obs},i)\,=\,w_i\,p_{\rm inf (i)}\,,
\ee
where $w_i$ is a weighting factor accounting for various further effects.
For example, it is intuitive that a larger number of observers living in $i$ results in a larger probability.\footnote{Various anthropic effects, most notably the number density of observers on the reheating surface, may be absorbed in $w_i$. Furthermore, $p({\rm obs},i)$ is proportional to the probability of {\it starting} inflation in inf$(i)$. This is different from the probability $p_{{\rm inf}(i)}$ of {\it finding} the universe in inf$(i)$. The latter is enhanced if a given inflationary plateau has a particularly long lifetime before reheating. This difference is a non-exponential effect which may also be absorbed in $w_i$.
We will return to this point in Sect.~\ref{sect_explicit_predictions}. \label{footnote_p_obs}} 
Eventually, our predictions are ratios
\begin{align}
    \frac{p({\rm obs},i)}{p({\rm obs},j)}=\frac{w_i\,p_{\rm inf (i)}}{w_j\,p_{\rm inf (j)}}\,,\label{anthr_prediction_w}
\end{align}
for any two anthropically viable vacua $i,j$ (i.e.~$w_i\neq 0\neq w_j$) of the landscape. To calculate the $p_{\rm inf (i)}$, we treat the inflationary plateaus of any scalar potential as additional dS vacua of the landscape. 
They will hence appear in the rate equation \eqref{rate_equation} with a total decay rate dominated by $\Gamma_{{{\rm inf}(i)}\to i}$ corresponding to inflation ending by reheating in vacuum $i$.
Solving the rate equations, which we will do explicitly in Sect.~\ref{sect_calculations}, yields the $p_{\rm inf (i)}$ as functions of all vacuum transition and creation rates.

Let us for the moment assume that observers can only live in universes which are asymptotically dS, i.e.~consider the situation where $i,j$ feature a (small) cosmological constant.
Since dS spaces are viewed from the perspective of the static patch, which is finite in size, only a finite number of observers will live there. 
As a result, the ratio $w_i/w_j$ is well defined and not exponentially large or small in the dS radii $\ell_{dS,i}$ and $\ell_{dS,j}$. This is clear since the size of the static patch is $\sim \ell_{dS}^3$. We then approximately have
\begin{align}
    \frac{p({\rm obs},i)}{p({\rm obs},j)}\simeq \frac{p_{\rm inf (i)}}{p_{\rm inf (j)}}\,.\label{anthr_prediction}
\end{align}

We conclude that under the assumption that observers can only live in dS vacua, the leading order prediction for the most likely anthropic vacuum is solely determined by the largest value of $p_{\rm inf (i)}$, which can be computed from \eqref{rate_equation}.

\section{Projecting on observers in the presence of terminal vacua}\label{sect_projecting_on_observers}
In the previous section, we have seen how to make predictions in case observers can only live in asymptotically dS vacua. In this section, we will discuss the general situation, allowing for observers in terminal vacua. 
This will lead to difficulties in particular in the case of observers in Minkowski space, to which we refer as the `Minkowski-space infinity problem'. Subsect.~\ref{sect_summary_on_counting} provides an executive summary of the main results of this discussion.

\subsection{Summary of approaches to the Minkowski-space infinity-problem}\label{sect_summary_on_counting}
\begin{figure}[ht]
	\centering
    \def\svgwidth{0.95\linewidth}
	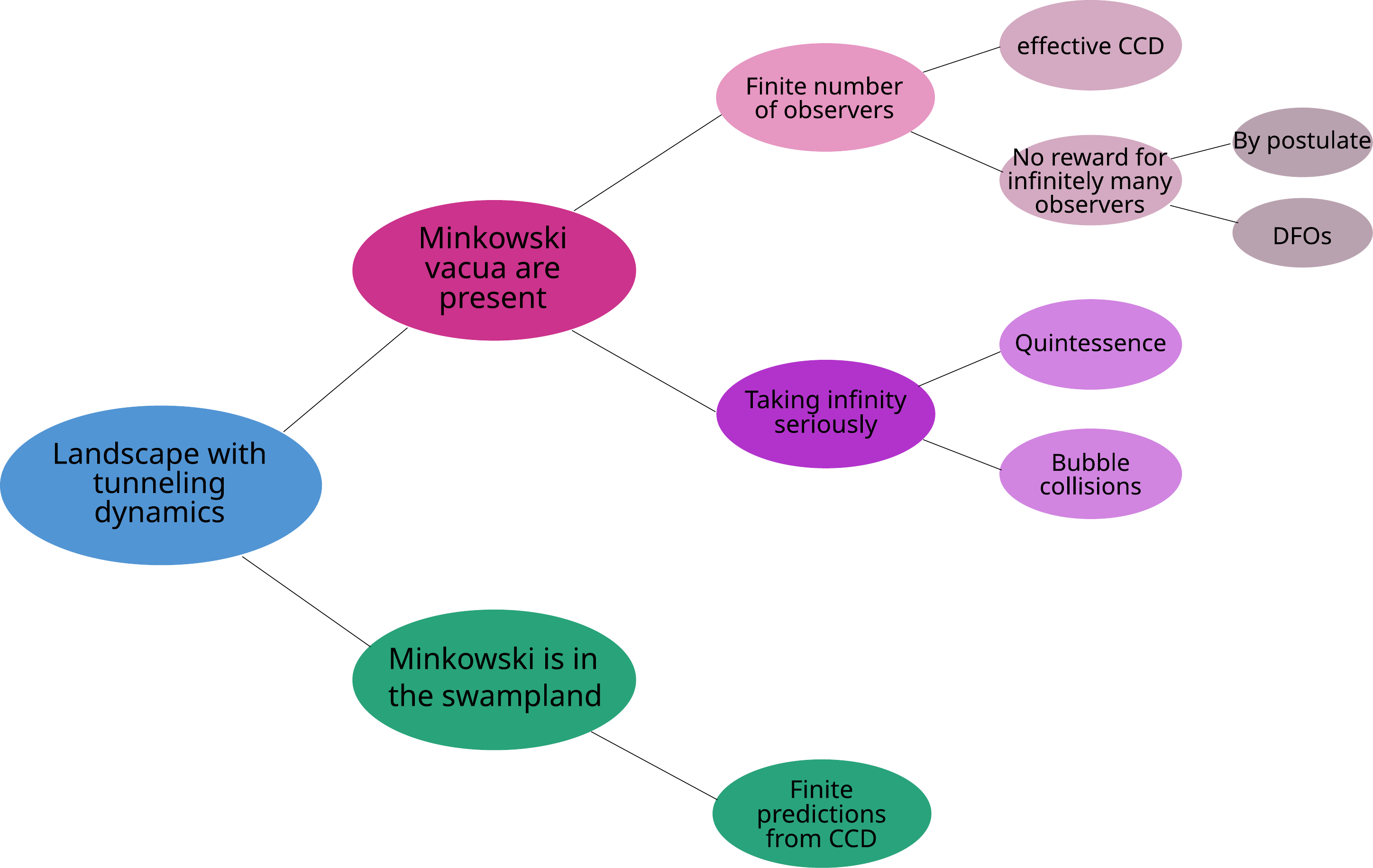
	\caption{Approaches to the Minkowski-space infinity problem.}
	\label{fig_tree_diagram}
\end{figure}
If observers can exist in AdS and Minkowski vacua, the finiteness argument based on the finite dimensionality of dS Hilbert spaces fails. In the case of AdS, this may be argued to be irrelevant since the unavoidable big crunch allows only finite portions of a reheating surface in AdS to be causally connected. For a reheating surface in Minkowski space, no such argument can be made. Moreover, even if no Minkowski-space reheating surfaces with observers exist, the Minkowski-space infinity problem still persists because of bubble collisions: In such collisions, infinite parts of dS-space reheating surfaces can make it into a  Minkowski region \cite{Kleban:2011pg}.

We explain this in more detail in Sects.~\ref{fbfh}-\ref{noet} and analyze possible resolutions. Let us here give a brief and systematic summary, following the diagram in Fig.~\ref{fig_tree_diagram}:

\begin{enumerate}
    \item \underline{\textbf{No Minkowski space:}}
        The issue disappears if Minkowski vacua are in the Swampland.
    \item \textbf{\underline{Finite number of observers in Minkowski space:}}
    Two lines of argument may avoid observer-infinities in Minkowski space:
    \begin{itemize}
        \item[2a] \underline{Effective CCD:} The CCD declares that only a finite portion of any spacelike surface in dS is physical. By analogy, we propose that only a finite portion of any reheating surface, including in Minkowski space, contains independent observers.
         \item[2b] \underline{No reward for infinitely many observers:}
         Infinite observer counts may be irrelevant for probabilistic predictions. This may either hold as a postulate \cite{Hertog:2013mra,Hartle:2016tpo} or because observers contain finitely many bits: They can then be grouped in a finite set of equivalence classes, which we call `dressed finite observers' or `DFOs'.
    \end{itemize}
    \item \textbf{\underline{Taking infinities seriously:}} If infinities of observers in Minkowski space do indeed affect predictions, this may happen in two distinct ways (or through a combination thereof):
       \begin{itemize}
        \item[3a] \underline{Predicting reheating to Minkowski:}
        If reheating to Minkowski is possible, then anthropic arguments predict Minkowski space in our asymptotic future, suggesting quintessence-like dark energy in our Universe.
         \item[3b] \underline{Prediction based on bubble collisions:}
         If there is no direct reheating to Minkowski, then anthropic arguments suggest that we will collide with a Minkowski bubble in the future, introducing a dependence on the dynamics of bubble collisions \cite{Chang:2007eq,Aguirre:2009ug,Freivogel:2011eg}.
    \end{itemize}
\end{enumerate}

The reader not satisfied with this short version may now read Sects.~\ref{fbfh}-\ref{noet} and, if desired, then return to the present subsection. We note in particular that Sect.~\ref{noet} provides a brief discussion of landscapes without eternal inflation, a subject we did not repeat in this summary.

\subsection{Finiteness based on a finite observer Hilbert space}
\label{fbfh}
In general, $\Psi$ lives in an infinite-dimensional Hilbert space and may hence not be normalizable. As a result, ratios of the form \eqref{Local_WDW_prediction} may be ill-defined. For example, $P_\alpha$ and $P_\beta$ could be projection operators on AdS vacua. Since the relevant Hilbert subspaces are infinite-dimensional, the numerator and denominator could both be infinite, such that their ratio cannot be quantified. For anthropic predictions, however, we are interested in evaluating \eqref{Local_WDW_prediction}
for operators $P_\alpha,P_\beta$ which involve a projection on the subspace $\mathcal{H}_{obs}\subset \mathcal{H}$, describing observers. This projection can restore normalizability, such that the corresponding predictions are well-defined.
If anthropic observers live only in dS universes, $\mathcal{H}_{obs}$ is contained in the finite-dimensional dS part of $\mathcal{H}$. The projections of $\Psi$ on $\mathcal{H}_{obs}$ or on subspaces thereof are then normalizable. This is the basic origin of our claim that the ratio $w_i/w_j$ in \eqref{anthr_prediction_w} is finite for $i,j$ denoting dS vacua.

\subsection{Finiteness lost}
\begin{figure}
	\centering
	\def\svgwidth{0.35\linewidth}
	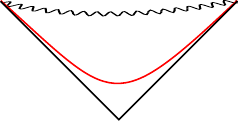
	\caption{An AdS bubble with crunching surface (wiggly line) and reheating surface (red line) is shown.}
	\label{fig_AdS_crunch}
\end{figure}

Unfortunately, the assumption that observers only live in dS is hard to justify. Indeed, certain inflationary plateaus may lead to reheating in AdS. While such an inflationary bubble will eventually end in a big crunch singularity, it will contain an infinite reheating surface (cf.~Fig.~\ref{fig_AdS_crunch}). Semiclassically, this surface may contain infinitely many observers. Together with the fact that the relevant part of $\Psi$ now clearly lives in a infinite-dimensional Hilbert space, we have lost our previous, simple argument for finiteness.

At this point, going beyond semiclassics, one may try to argue that the presence of the big crunch singularity somehow limits the number of observers that can be considered independent. For example, there are only finitely many observers in any causally connected region of the spacetime displayed in Fig.~\ref{fig_AdS_crunch}.

However, such a reasoning cannot be applied if reheating occurs in a Minkowski vacuum. To save finiteness, one may appeal to the very reasonable expectation that Minkowski vacua are always $\mathcal{N}=2$ supersymmetric as otherwise corrections to the scalar potential will induce a non-zero vacuum energy.
Since $\mathcal{N}=2$ SUSY excludes chiral matter, it might become impossible to form observers (see e.g.~\cite{Douglas:2012bu} for a similar argument).

\begin{figure}
	\centering
    \def\svgwidth{0.99\linewidth}
	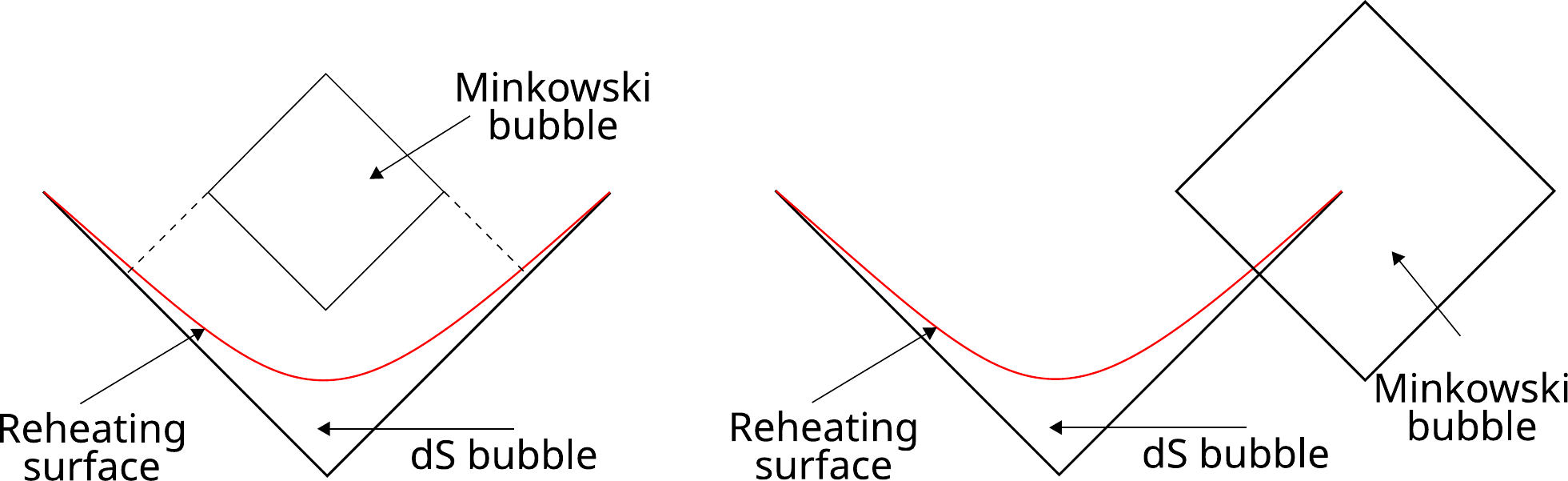
	\caption{Illustration of a Minkowski bubble inside a dS bubble (left) and a colliding bubble system (right). On the left, the part of the reheating surface that can send signals into the bubble is finite, meanwhile on the right it is infinite.}%
	\label{fig_bubbles}
\end{figure}

Even if such an argument were valid, we are forced to allow for situations where an inflationary dS bubble with observers on the reheating surface and an additional Minkowski bubble are both present (cf.~Fig.~\ref{fig_bubbles}).
A Minkowski bubble that forms completely inside an inflationary dS bubble still does not feature infinitely many observers from the reheating surface since the region of the reheating surface being in causal contact with the Minkowski bubble is only finite in size.
The situation is different for a Minkowski bubble colliding with a dS bubble, where the intersection of the reheating surface with the colliding bubble is an infinite 2d surface, cf.~\cite{Kleban:2011pg}.
We thus face the unavoidable situation that infinitely many observers (or at least their remnants) enter a Minkowski bubble. Now the CCD can provide no straightforward argument why only a finite region of the reheating surface should be considered.\footnote{A similar observation has already been made in \cite{Arkani-Hamed:2007ryv}}.
In other words, it appears as if there are vacua in the landscape contributing to \eqref{anthr_prediction_w} with an infinitely large anthropic weight $w$.

\subsection{Finiteness regained}\label{sect_finiteness_regained}
\paragraph{Effective CCD:} In the semiclassical picture of the multiverse, the reheating surfaces in nucleated bubbles are infinitely large. As a result, infinitely many observers will form on a given reheating surface. If the considered pocket universe is dS, the CCD implies that the infinity is an overcounting of the same information and only finitely many observers should be considered independent: The CCD limits the counting of observers to a single static patch.

\begin{figure}
	\centering
	\def\svgwidth{0.8\linewidth}
	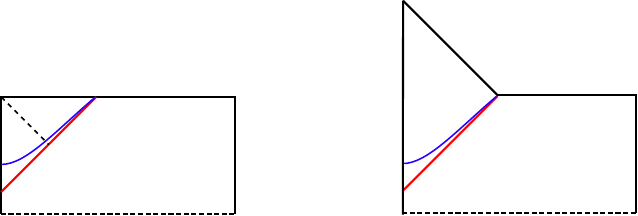
	\caption{The creation of a spherical universe with a subsequent formation of a dS bubble (left) and Minkowski bubble (right) are displayed. The red line denotes the bubble wall and the blue line illustrates the location of the reheating surface. The diagonal dashed line on the left shows the cosmological horizon for an observer located on the vertical solid black line on the far left.}
	\label{fig_bubbles_reheating}
\end{figure}

At first sight, the situation appears to be qualitatively different when considering a universe with inflation and reheating to Minkowski space -- the two cases are compared in Fig.~\ref{fig_bubbles_reheating}.
Nevertheless, the observers living on the reheating surfaces experience in both situations a certain energy density which, in case of dS observers, contains a contribution from the cosmological constant.
However, if the cosmological constant only provides a negligible contribution to the energy budget, there is no observable difference between the two reheating surfaces in Fig.~\ref{fig_bubbles_reheating}.
Such a difference only becomes relevant in the distant future, far away from the reheating surface (where all observers may have died).
Thus, given that only finitely many observers are independent in the dS case, it appears counter-intuitive that the subdominant effect of the cosmological constant makes a qualitative difference in the counting of independent observers.
This suggests that one should reconsider what the proper quantum gravity way of counting observers on reheating surfaces of Minkowski bubbles is. It should not be taken for granted that the result is infinite. Instead, we should be looking for a way of counting that is similar to the counting of observers living in dS bubbles with a small cosmological constant. 

Thus, let us take a fresh look at how many independent observers there are on a reheating surface in Minkowski space: Consider the nucleation of an inflationary bubble, i.e.~a bubble in which a scalar is initially rolling along an inflationary plateau. We view this as an approximate dS space with entropy $S_{\rm inf}$. The initial quantum state after nucleation is contained in an $\exp(S_{\rm inf})$-dimensional subspace of $\mathcal{H}$.  At late times, the state explores the entire infinite-dimensional Hilbert space associated to Minkowski space and one can imagine that there is a continuous transition letting the dimension of the accessible Hilbert space grow during the cosmological evolution.
One may then speculate that, at the moment of reheating, the dimension of the relevant Hilbert space is set by an entropy $S_{\rm reh}\sim M_P^2/H_{\rm reh}^2$, with $H_{\rm reh}$ the Hubble scale on the reheating surface. For simplicity, we do not distinguish between the moment of reheating and the moment of star- and observer formation. We hence identify $H_{\rm reh}$ with the curvature scale relevant for the formation of observers.
By analogy to the dS situation, we assume that the Hilbert space of dimension $\exp(S_{\rm reh})$ describes a region of size $1/H_{\rm reh}^3$, containing the independent observers.\footnote{One may think of the CCD as the statement that there exist gauge redundancies along surfaces of constant field value (i.e.~on slices of constant FLRW time).} As a result, ratios of anthropic weighting factors $w_i/w_j$ are finite and not exponentially large or small.
Such a reasoning can be viewed as an effective interpretation of the CCD, where passing through a nearly-de Sitter state suffices for making the accessible Hilbert space finite.\footnote{In the context of the census taker and causal diamond measure, it was speculated \cite{Bousso:2011up} on the basis of holographic arguments  \cite{Freivogel:2006xu} that information in Minkowski bubbles is redundant.} Clearly, the same finiteness argument also applies to the case when reheating to AdS is considered.

This argument resonates nicely with the finding \cite{Arkani-Hamed:2007ryv} that requiring validity of effective field theory combined with the CCD limits the number of independent modes of inflationary origin. If for example our Universe is rolling into a Minkowski vacuum without cosmological horizon, the number of fluctuations originating from inflation that can enter our horizon is in principle unbounded. However, the CCD implies that these modes must be correlated, leaving at most $\order{S_{\rm inf}}$ many independent degrees of freedom, with $S_{\rm inf}$ denoting the entropy of the approximate dS universe during inflation. This situation is fully analogous to radiation coming out of a black hole during the evaporation process which at late times must be correlated with the degrees of freedom already being outside the horizon \cite{Page:1993wv}. The statement of the CCD that $\dim{\mathcal{H}_{dS}}<\infty$ is then analogous to the claim that the transition of modes into Minkowski space follows similar principles as the process of degrees of freedom leaving the black hole horizon. 

\paragraph{Equivalent observers:}
Independently of the above, there is a further way of reasoning why well-defined predictions can be derived in the presence of Minkowski vacua: One may argue that anthropic observers necessarily correspond to finite subsystems of the total space accessible to the wave function of the universe. In other words, they have a finite-dimensional Hilbert space, $\dim\,{\cal H}_{obs}<\infty$. By virtue of this definition, any observer can encode only a finite amount of information about the state of the rest of the universe (roughly $S_{obs}$ bits, with the finite-dimensional observer Hilbert space allowing for about $e^{S_{obs}}$ different inequivalent quantum states which the observer can be in). Therefore, the future-infinite amount of observers on open slices of Minkowski spacetime inside the bubble will decompose into a finite ${\mathcal O}(e^{S_{obs}})$ set of equivalence classes which we may call `dressed finite observers' (DFOs). In the limit to asymptotic future, each DFO contains infinitely many finite observers in the same observer quantum state.  

We can now declare that a `real' observer must be identified with a DFO since the infinitely many observers in that DFO equivalence class perceive themselves and their internal description of the rest of the universe to be exactly the same, and are thus inseparable by any operation accessible for a finite observer. This yields a finite observer count even in Minkowski space. We note that there is no intrinsic cutoff on the size of the finite-dimensional Hilbert space and thus $S_{obs}$ per se. In this sense, there is a minimal amount of `anthropic' modeling in our choice of DFOs, in that we manifestly limit $\dim\,{\cal H}_{obs}<\infty$ and $S_{obs}<\infty$ to be small in a sense of being parametrically tiny compared to the dS horizon entropies of cosmological constants corresponding to EFT-accessible sub-Planckian vacuum energy densities. This notion thus is guided by the smallness of the entropy of human-like observers relative to the late-time dS entropy of our observable universe.

Furthermore, it was claimed elsewhere that one should not focus on the total number of observers in a given universe \cite{Hertog:2013mra,Hartle:2016tpo}. Instead, one should demand that the probability for at least one observer forming is near unity.
Given this proposal, an infinity of observers that can exist in a Minkowski bubble poses no threat to deriving anthropic predictions from the local WDW-measure. 

\paragraph{No Minkowski space:} A particularly drastic possibility for how infinities associated with Minkowski space could be avoided is the following: Our full quantum gravity theory might simply not contain Minkowski vacua. Instead, there could be only (metastable) dS states and AdS vacua. This resonates with the `fundamentalist-AdS/CFT' point of view that the only way to define quantum gravity is via a CFT. Presumably metastable dS or slow-roll states can then only arise as excitations of AdS (cf.~also~\cite{Freivogel:2004rd, Maltz:2016iaw,Dvali:2017eba}).

\paragraph{Taking infinitely many Minkowski-space observers seriously:} By contrast, we can also not exclude the possibility that infinite observer counts in Minkowski space are indeed physical. One can argue for this by including cosmological Minkowski vacua in our WDW approach on the basis of a `FRW-slicing', with each slice being a hyperboloid of constant density. It could then be that no variant of the CCD applies, i.e.~there are no gauge redundancies along each surface. This clearly leads to infinities of observers. Alternatively, one could slice the Minkowski space on the r.h.~side of Fig.~\ref{fig_bubbles_reheating} horizontally, such that each slice has only a finite number of observers. Nevertheless, infinities can still arise because these slices continue to intersect the reheating surface all the way into the infinite future. In fact, these intersection loci keep growing.
In any case, once we accept these infinities as real, one straightforwardly concludes that our measure prefers universes in which our FRW-cosmology asymptotes to Minkowski space. In other words, the presently observed dark energy should decay, possibly in a quintessence-like manner. However, this conclusion might be premature: Indeed, we might live in a landscape where no observers form on reheating surfaces leading to Minkowski space. This could come about because Minkowski space requires ${\cal N}=2$ SUSY or SUSY-restoration through decompactification. In such a landscape, the preferred vacua would be those which have a large probability of colliding with Minkowski bubbles. This is similar to the predictions obtained from the census taker measure, cf.~\cite{Freivogel:2011eg,Chang:2007eq,Aguirre:2009ug}.

\subsection{A landscape without eternal inflation}
\label{noet}
It has been suggested that, while being compatible with slow-roll inflation, quantum gravity does not allow for de Sitter minima and eternal inflation because of a fundamental inconsistency caused by quantum breaking \cite{Dvali:2013eja, Dvali:2014gua, Dvali:2017eba, Dvali:2018fqu, Dvali:2018jhn, Dvali:2020etd, Dvali:2021kxt}. This is also a possible implication of the de Sitter conjecture in string theory \cite{Danielsson:2018ztv, Obied:2018sgi, Garg:2018reu, Ooguri:2018wrx}. If so, creation from nothing can only lead to states on an inflationary slow-roll plateau, with further tunneling processes not playing a significant role.
If furthermore bubbles of something \cite{Hawking:1998bn, Turok:1998he, Garriga:1998ri, Bousso:1998pk, Blanco-Pillado:2011fcm, Friedrich:2023tid} do not exist, the closed topology of the dS universes created according to Hartle-Hawking or Linde/Vilenkin ensures finiteness of the reheating surface and the number of observers.
As a result, in such a landscape no CCD-based argument is needed for making anthropic predictions. If, however, bubbles of something can occur, one nevertheless has to deal with infinite reheating surfaces. The arguments of Sect.~\ref{sect_summary_on_counting} for treating infinities are then relevant even though eternal inflation is not possible. A different way in which infinite reheating surfaces can arise in the absence of metastable de Sitter is through topological inflation \cite{Linde:1994hy, Vilenkin:1994pv}. In this context, it is an interesting question to what extent arguments against de Sitter or other swampland criteria also affect topological inflation \cite{Hebecker:2017wsu, Dolan:2017vmn,
Dvali:2018jhn, Cai:2019dzj, Lin:2019fdk, Fanaras:2023acz}.

\section{Explicit predictions in the multiverse -- basic ingredients}\label{sect_strat}

Let us now try to make progress towards explicit prediction in the multiverse using the local WDW measure.
We assume that eternal inflation is not excluded and that the finiteness arguments of Sect.~\ref{sect_finiteness_regained} hold, such that anthropic predictions are not a priori dominated by Minkowski universes. This corresponds to the cases 1.~and 2.~of Sect.~\ref{sect_summary_on_counting}. 

More precisely, we base our further analysis on \eqref{anthr_prediction} and it is crucial for us that the transition from \eqref{anthr_prediction_w} to \eqref{anthr_prediction} still holds.
This is obvious in case 1.~of Sect.~\ref{sect_summary_on_counting}. In case 2.a, it holds because ratios $w_i/w_j\propto \ell_{reh,i}^3/\ell_{reh,j}^3$ are not exponentially large. In case 2.b, it holds either by definition or, if we rely on the DFO argument, using the extra assumption that the number of independent DFOs is not exponentially large. In any case, evaluating \eqref{anthr_prediction} will now be our main focus.

\subsection{Solving the rate equation}\label{sect_calculations}
Our first goal is to solve \eqref{rate_equation} for the $p_i$. Since the normalization of creation rates is unknown, the $p_i$ can only be determined up to a multiplicative constant. This is fine since predictions are of the form $p_i/p_j$.
Equation \eqref{rate_equation} can be written in matrix form,
\begin{align}
    J_i=M_{ij}p_j\,, \quad M_{ij}=\delta_{ij}\Gamma_i-\Gamma_{j\to i}\,,\label{transition_matrix}
\end{align}
with $\Gamma_i$ being the total decay rate of vacuum $i$.
This equation can be solved in the form of the series (see Appendix \ref{appendix_solving_rate_equation} for more details):
\begin{align}
    p_i = \sum_{N=0}^\infty p_{i,N}\,, \quad p_{i,0}=\frac{J_i}{\Gamma_i}\,,\quad	p_{i,N}=\frac{1}{\Gamma_i} \sum_{j_1,...,j_N}J_{j_1}\frac{\Gamma_{j_1\to j_2}}{\Gamma_{j_1}}\frac{\Gamma_{j_2\to j_3}}{\Gamma_{j_2}}\hdots \frac{\Gamma_{j_N\to i}}{\Gamma_{j_N}}\,.\label{p_i_solution}
\end{align}
We note that the structure of \eqref{p_i_solution} is straightforward to interpret. In the sum over $N$, the $N^{\text{th}}$ term corresponds to $N$ tunneling processes between creation from nothing and vacuum $i$. Each tunneling process is suppressed by a small number coming from a `branching ratio' of the form $\Gamma_{j_s\to j_{s+1}}/\Gamma_{j_s}\ll 1$.
It becomes apparent that we need to investigate both the creation rates $J_i$ as well as the tunneling rates $\Gamma_{i\to j}$ in more detail.

\subsection{Creation rates}\label{sect_creation}
The traditional proposals for the creation of universes from nothing are the Hartle-Hawking no boundary \cite{hartle1983wave} and the Linde/Vilenkin tunneling \cite{Linde:1983mx,Vilenkin:1984wp} proposals. 
Both study the quantum mechanical 
nucleation of compact spherical universes which eventually go on shell and expand.
In both cases, the basic ingredient is an instanton geometry which describes euclidean dS, i.e.~a 4-sphere. Its action $\mathcal{S}$ is given by minus the dS entropy, $\mathcal S=-S_{ds}$, with
\begin{align}
S_{dS}=24\pi^2M_P^4/V_{dS}=8\pi^2M_P^2\ell_{dS}^2\,.\label{S_Entropy_form}
\end{align}
Here $V_{dS}$ is the energy density and $\ell_{dS}$ the dS radius. The proposed rates are
\begin{align}
    J\propto \begin{cases} \exp(S_{dS}) & \text{Hartle-Hawking \cite{Hawking:1998bn}}\\ 
        \exp(-S_{dS}) & \text{Linde/Vilenkin \cite{Linde:1983mx, Vilenkin:1984wp}}
        \end{cases}\, .\label{source_strength}
    \end{align}
Clearly Hartle-Hawking favor small and Linde/Vilenkin favor large vacuum energies. 

A different proposal for creating universes from nothing was introduced in \cite{Hawking:1998bn,Turok:1998he} and further developed in \cite{Garriga:1998ri,Bousso:1998pk,Blanco-Pillado:2011fcm}. Its key ingredient are end-of-the-world (ETW) branes, representing boundaries of 4d spacetime. The existence of such boundaries allows for the nucleation of universes with the geometry of a ball, bounded by a spherical ETW brane.
The prime example of an ETW brane arises in 5d to 4d compactifications, where the $S^1$ compact space can shrink to zero size at a codimension-one locus over the 4d base.
This gives rise to Witten's famous `bubble of nothing' \cite{Witten:1981gj}.
The creation process we are discussing may be thought of as the inverse process, i.e.~a `bubble of something' \cite{Friedrich:2023tid}.\footnote{In \cite{Blanco-Pillado:2011fcm}, the term `bubble from nothing' was used instead.}
The relevance of this creation process is enhanced by the cobordism conjecture \cite{McNamara:2019rup}, implying that every vacuum of a complete theory of quantum gravity exhibits ETW branes.
Further work on ETW branes include \cite{Angius:2022aeq,Blumenhagen:2022mqw,Sugimoto:2023oul,Muntz:2024joq}.
An explicit geometry of such ETW branes in the type IIB flux landscape was provided in \cite{Friedrich:2023tid}, where first steps towards an analysis of the corresponding creation process in string theory were also made. As in the no boundary proposal, ball-shaped universes nucleate quantum-mechanically,  go on-shell and subsequently expand. By contrast, however, this process also allows for the creation of Minkowski and AdS vacua.
The bubble-of-something process produces open Friedmann universes, similar to the interiors of Coleman-de Luccia (CdL) bubbles \cite{coleman1980gravitational}.

Recall that we are only interested in observers on post-inflationary reheating surfaces and that we treat inflationary plateaus as short-lived dS vacua. We may hence restrict our attention to the bubble-of-something creation of dS.
In the regime of a thin ETW brane, the relevant euclidean instanton action reads \cite{Friedrich:2023tid}
\begin{align}
    \mathcal{S}=-4\pi^2M_P^2\ell_{dS}^2\left(1\pm\sqrt{\frac{T_4^2\ell_{dS}^2}{T_4^2\ell_{dS}^2+4M_P^4}}\right)\,.\label{S_bos}
\end{align}
Here the upper sign is relevant for negative tension, $T_4<0$, and the lower sign for $T_4>0$.
Using \eqref{S_Entropy_form}, it is straightforward to see that $\mathcal{S}>-S_{dS}$, with equality holding in the limit $T_4\to -\infty$.\footnote{A bubble of something may be thought of as the creation of a universe together with a bubble of nothing. In the limit $T_4\to -\infty$, the initial size of the bubble of nothing vanishes.
In our understanding, the bubble of something in the limit $T_4\to -\infty$ may correspond to the `creation from nothingness' considered in \cite{Cespedes:2023jdk}.}
If one uses the `Hartle-Hawking sign-choice', i.e.~$J\sim \exp(-\mathcal{S})$, it then follows that the standard no-boundary-process is more likely than the bubble-of-something process. By contrast, adopting the `Linde/Vilenkin sign-choice', i.e.~$J\sim \exp(\mathcal{S})$, the bubble-of-something process wins. Specifically, the smallest enhancement arises if $T_4<0$ and $T_4^2\ell_{dS}^2\gg M_P^4$, in which case we have
\be
\mathcal S\simeq -S_{dS}+8\pi^2\frac{M_P^6}{T_4^2}\,.
\ee
The relative enhancement of the bubble-of-something relative to the Linde/Vilenkin process is then by
\be
\exp(8\pi^2 M_P^6 / T_4^2)\,,
\ee
which is enormous whenever the brane tension is sub-Planckian. In all other cases, the enhancement is by an exponent of the order of $S_{dS}$, i.e. even stronger. Thus, if the Linde/Vilenkin sign choice is the right one and ETW branes exist in the validity range of the EFT, the bubble-of something process is always completely dominant.
The strongest enhancement occurs for $T_4\ell_{dS}\gg M_P^2$, in which case 
\begin{align}
    \mathcal{S}\simeq -8\pi^2\frac{M_P^6}{T_4^2}\,.\label{large_T}
\end{align}

Finally, another possibility to create de Sitter vacua from nothing, again making use of the existence of ETW branes, was suggested in \cite{Friedrich:2024aad}. Analogously to the `No Boundary Proposal', this `Boundary Proposal' is based on an off-shell spherical universe in our past. However, instead of starting at zero size, our spatial 3-sphere starts its existence at a spacelike negative-tension ETW brane. 
The euclidean action for the corresponding instanton reads
\begin{align}
    \mathcal{S}=-8\pi^2M_P^2\ell_{dS}^2\sqrt{\frac{T_4^2\ell_{dS}^2}{T_4^2\ell_{dS}^2+4M_P^4}}\,.\label{S_boundary}
\end{align}
The boundary instanton again satisfies $\mathcal{S}>-S_{dS}$. It is hence again suppressed relative to the standard no-boundary process in case of the Hartle-Hawking sign choice and enhanced in case of the Linde/Vilenkin sign choice. As before, the smallest possible enhancement occurs for $T_4<0$ and $T_4^2\ell_{dS}^2\gg M_P^4$, in which case the exponent is
\begin{align}
    \mathcal{S}\simeq -S_{dS}+16\pi^2\frac{M_P^6}{T_4^2}\,.
\end{align}
Thus, whenever the EFT is valid, the Boundary Proposal rate completely dominates over the Linde-Vilenkin rate.

The enhancement is the strongest in the limit $|T_4|\ell_{dS}\ll M_P^2$, in which case we find
\begin{align}
    \mathcal{S}=-4\pi^2|T_4|\ell_{dS}^3\,.\label{S_bdry_small_T}
\end{align}
While the exponential suppression completely disappears for $T_4\ll \ell_{dS}^{-3}$, this has to be taken with some caution: In the formulae above, non-exponential prefactors have remained undetermined, so more research is needed to quantify the outcome of this proposal in the limit of small $T_4$.

\subsection{Nontrivial topology}
\label{topsec}
The Linde/Vilenkin proposal for the creation of universes from nothing can be viewed as the statement that small spheres can quantum mechanically fluctuate from nothing and then undergo a tunneling process through a potential barrier to become on-shell. 
One manifestation of this interpretation is the fact that the Linde/Vilenkin wavefunction $\Psi_{LV}$ satisfies an inhomogeneous WDW equation with a localized source for zero-size universes, see e.g.~\cite{Feldbrugge:2017kzv}.\footnote{By contrast, the Hartle-Hawking wavefunction is real and may be viewed as a standing wave, not requiring any source \cite{Feldbrugge:2017kzv}.}
The potential barrier arises from the positive spatial curvature of the spherical universe.

Now, as observed in \cite{Zeldovich:1984vk,Coule:1999wg,Linde:2004nz}, if spherical universes at vanishing size can fluctuate from nothing, it appears inevitable that compact universes with different topology, e.g. small tori, can also nucleate at little to no cost. Since there is no spatial curvature, there is no potential barrier and hence no tunneling suppression for going on-shell. As a result, the creation of a toroidal or negatively curved compact universe appears to be far more likely than the creation of ordinary de Sitter space with spherical spatial section.  Moreover, there is no exponential preference for a large cosmological constant of the nucleated vacuum.\footnote{It was argued in \cite{Rubakov:1984bh,Rubakov:1984ki} that the creation of spherical universes may, in fact, also be free of any tunneling suppression. The reason is that, given the Linde/Vilenkin sign choice, fluctuations are enhanced and these compensate the curvature term in the action which was responsible for the suppression.}
It was speculated that subleading effects such as Casimir energy \cite{Zeldovich:1984vk} or the gradient energies of scalar fields \cite{Linde:2004nz} determine the most likely properties of the created universe. 

Regarding applications to the string landscape, one might expect that also details of the compactification, e.g.~the geometry of the internal manifold, affect which vacua are preferred in the creation process. 
Work in this direction includes \cite{Conti:2014uda,Lehners:2022mbd,Hertog:2021jyd,Fanaras:2021awm,Fanaras:2022twv}. 
However, in the present analysis we adopt a 4d EFT approach and hence do not account for such effects. This amounts to assigning all dS universes which avoid a potential barrier by their nontrivial topology a similar probability for creation from nothing.

\subsection{Tunneling rates}
In the thin wall limit, tunneling rates are determined by the domain wall tension $T$, by the true- and false-vacuum energy densities $V_t,V_f$, and the Planck mass $M_P^2=1/(8\pi G)$\cite{coleman1980gravitational}. 
The false-vacuum decay rate $\Gamma$ is conveniently expressed using the dimensionless variables \cite{coleman1980gravitational,Parke:1982pm}
\begin{align}
        x=\frac{3T^2}{4M_P^2(V_f-V_t)}\,,\qquad y=\frac{V_f+V_t}{V_f-V_t}\,,\qquad r(x,y)=2\frac{1+xy-\sqrt{1+2xy+x^2}}{x^2(y^2-1)\sqrt{1+2xy+x^2}}\,.\label{dimensionless_variables}
\end{align}
It reads
\begin{align}
    \Gamma \sim \exp(-B)\,\,,\qquad\mbox{with}\qquad B=\frac{27\pi^2T^4}{2(V_f-V_t)^3}r(x,y)\,.\label{decay_rate}
\end{align}
The importance of gravitational effects is measured by $x$. In the regime $x\ll 1$, $xy\ll 1$, the expression for the decay rate reduces to the field theory result, given by $r(x,y)\to 1$.
If $x$ or $xy$ become $\order{1}$ or larger, gravitational effects are important and are quantified by the function $r(x,y)$.
In the limit $x\gg1$, $x\gg y$, the tunneling exponent $B$ approaches the entropy $S_f$ of the decaying vacuum. Hence, the decay occurs on the recurrence time scale of the corresponding dS space. See App.~\ref{Appendix_rate_discussion} for some further details and useful formulae.

\section{Tunneling rates in the string landscape}\label{sect_String_rates}
In this section, we study tunneling rates for the most promising dS vacuum candidates in the string landscape: Type IIB flux compactifications on Calabi-Yau orientifolds with Kahler moduli stabilization \`{a} la KKLT \cite{Kachru:2003aw} or LVS \cite{Balasubramanian:2005zx}.
Some of the following results have appeared elsewhere in the literature.
Our summary of the various rates can be of general interest in studies of the string landscape.

We will show that, to the best of our present understanding, the dominant decay process of KKLT/LVS-type vacua is the destruction of the uplift.  
Tunneling between two dS vacua occurs only via flux transitions and is generically close to being maximally suppressed.
While tunneling to decompactification is generically slower than the destruction of the uplift, it still proceeds much faster than tunneling between dS vacua.

\subsection{Tunneling to decompactification} \label{ssec:decompactification}
This tunneling transition always exists and its final state is 10d Minkowski space. The tension of the domain wall has been estimated for both KKLT and LVS models \cite{Westphal:2007xd}. For generic vacua, the resulting decay exponent $B$ 
is given by 
\begin{align}
    B_{\rm dec}=\frac{S_f}{\left(1+1/x\right)^2}\qquad \text{ with }\qquad x\lesssim 1\,,\label{B_dec}
\end{align}
and is thus smaller but of a similar order of magnitude as the dS entropy $S_f$. Details of the analysis are given in Appendix \ref{appendix_sect_tunneling_calculations}. If, however, the vacuum energy of the dS space is fine-tuned to be very close to zero, the decay exponent becomes even larger and approaches $S_f$ \cite{Kachru:2003aw,Westphal:2007xd}.

\subsection{Flux transitions} \label{ssec:flux_transition}
Flux transitions are mediated by domain walls made from D5/NS5 branes wrapped on CY 3-cycles. They can always occur and their final state can be either a dS or an AdS vacuum. The latter case will not be discussed here since, as will become clear in the next subsection, fast transitions to AdS always exist due to the `destruction of the uplift'. Thus, the decay to AdS via flux transitions is not essential.

Generically, both transitions from KKLT to KKLT as well as from LVS to LVS occur in a regime where gravitational effects are large. In both cases, the decay exponent can be approximated as
\begin{align}
	B_{\rm flux}=S_f-\frac{64\pi^2M_P^6}{T^2}\label{B_flux}\,,
\end{align}
with $64\pi^2 M_P^6/T^2\ll S_f$. This is shown in App.~\ref{appendix_sect_tunneling_calculations}, which in part reviews the analysis of \cite{Kachru:2002ns,deAlwis:2013gka}. It is easy to convince oneself that this result also holds for transitions from KKLT to LVS and vice versa.
By comparing with \eqref{B_dec}, we notice that generic flux transitions are significantly suppressed compared to tunneling to decompactification.

In \cite{Aguirre:2009tp}, the backreaction of the D5/NS5-brane domain walls on the volume has been studied.  It was found that for generic transitions in the KKLT and LVS landscape, the backreaction is relevant and it has even been questioned whether flux transitions can occur at all.
In \cite{Brown:2011ry} it was argued that non-instantonic transitions are nevertheless expected as long as there is no symmetry forbidding them.  Decay rates for such transitions were estimated in \cite{Blanco-Pillado:2019xny} in toy-landscapes.
For situations where also a CdL transition exists, it was found that the non-instantonic decay exponent for down tunneling is slightly larger than the corresponding CdL exponent.
In the case of downtunneling-events, Eq.~\eqref{B_flux} therefore both represents a lower bound on the decay exponent and may be viewed as a reasonable approximation so that Eq.~\eqref{B_flux} can serve as an order-of-magnitude estimate even if instanton transitions do not exist. However, for up-tunneling it was argued that the decay exponent can be significantly smaller than its CdL analogon.

Finally, non-generic flux transitions featuring a much smaller decay exponent than \eqref{B_flux} could exist. Specifically, if the domain wall tension $T$ can be tuned very small, the decay occurs in the field-theoretic regime. 
In principle, a lower bound on the domain wall tension follows from a BPS argument. 
Although SUSY is broken in our cases of interest, branes can nevertheless be `near BPS' with respect to the SUSY theory of the CY before fluxes and uplifting.
As explained in \cite{Lust:2022lfc}, BPS domain walls cannot arise from wrapping a brane on an arbitrarily small cycle without the loss of geometric control.
As a result, typical `near BPS' domain walls will not have a tension small enough to lead to a decay in the field-theoretic regime. 
However, controlled small-tension domain walls can nevertheless arise due to warping, such as in the presence of Klebanov-Strassler throats. 
Flux transitions in this region are then similar to the KPV decay \cite{Kachru:2002gs}, 
with a D5/NS5 wrapped on the $A$-cycle of a KS throat (cf.~Sect.~\ref{ssec:uplift}). Clearly, we also have to avoid producing too many D3 branes in such a warped-down flux transition, such that any pre-existing $\overline{D3}$ uplift in a different throat does not get destroyed. In summery, while they are highly non-generic and probably rare, we have a priori no reason to exclude such transitions altogether.

\subsection{The destruction of the uplift} \label{ssec:uplift}
Both KKLT and LVS models realize dS by uplifting an AdS minimum. The most-studied concrete realization is the $\overline{D3}$ uplift, which can decay by the KPV transition \cite{Kachru:2002gs}. This process, in which the $\overline{D3}$ brane(s) annihilates against flux, generally occurs in the field-theoretic regime with decay exponent \cite{Kachru:2002gs, Freivogel:2008wm} (cf.~App.~\ref{appendix_sect_tunneling_calculations})
\begin{align}
	B_{\rm KPV}= \frac{27b^6M^6g_s}{2048\pi (N_{\overline{D3}})^3}\,.\label{B_KPV_main}
\end{align}
Here $b\approx 0.932$, $M$ is a flux number, $g_s$ the string coupling and $N_{\overline{D3}}$ the number of decaying $\overline{D3}$ branes.

It is clear that the decay exponent \eqref{B_KPV_main} will very generally not be as large as for flux or decompactification transitions. Nevertheless, choosing appropriate values for $g_s, M$ and $N_{\overline{D3}}$ appears to allow for $B_{\rm KPV}\gg 1$ and hence long-lived dS. However, recent work shows that realizing such a parameter choice is non-trivial -- both in the KKLT \cite{Carta:2019rhx,Gao:2020xqh,McAllister:2024lnt} and the LVS \cite{Junghans:2022exo, Gao:2022fdi, Junghans:2022kxg, Hebecker:2022zme, Schreyer:2022len, Schreyer:2024pml} case: One the one hand, singularity and tadpole constraints limit the value of $M$. On the other hand, at small $M$ the KPV process is subject to curvature corrections to the action of the NS5 brane which mediates the decay \cite{Hebecker:2022zme, Schreyer:2022len, Schreyer:2024pml}. Thus, if a metastable $\overline{D3}$-uplift can be achieved at all, we expect settings where $B_{\rm KPV}$ is relatively small (maybe outside the regime where the derivation of \eqref{B_KPV_main} is reliable) to be more common. This strengthens our point that KPV transitions occurs much faster than decompactification or generic flux decays. As discussed in the last subsection, a non-generic flux transition could be similarly likely.

Let us briefly comment on the possibility that a KPV decay takes us from dS to dS: In principle, this is possible if there are two uplifts realized in two throats. However, the $\overline{D3}$ decay in one throat produces many D-branes which migrate and are very likely to annihilate all the $\overline{D3}$s of the other throat. Thus, following \cite{Frey:2003dm}, we conclude that KPV processes always or almost always end in AdS.

Of course, we should also consider different uplifting proposals, such as the $F$-term \cite{Saltman:2004sn} or T-brane \cite{Cicoli:2015ylx} uplift (for recent work on the former see \cite{Hebecker:2020ejb, Krippendorf:2023idy}). To the best of our understanding, all ideas of this type share the feature of a metastable SUSY-breaking minimum protected by a small barrier, such that a relatively low-energy process can return the model to a non-uplifted AdS phase. We then expect that also the decay of these alternative uplifts proceeds in the field-theory regime and hence much faster than flux or decompactification decays.

\subsection{Bubbles of nothing}
A bubble of nothing is a vacuum decay process discovered in \cite{Witten:1981gj} (cf.~\cite{Blanco-Pillado:2011fcm,Blanco-Pillado:2016xvf,GarciaEtxebarria:2020xsr,Draper:2021qtc,Bomans:2021ara,Angius:2022aeq} for a selection of recent work). In the simplest case, the underlying geometry is $\mathbb M_4\times S^1$, i.e.~4d flat space with one internal dimension. The relevant instanton is obtained by letting the internal $S^1$ shrink to zero size over the locus of an $S^3$ in the non-compact space. From the 4d perspective, this locus is an ETW brane with negative tension (the 4d EFT description was in particular developed in~\cite{Friedrich:2023tid}).
In the Lorentzian evolution, this ETW brane expands and eats up all space. According to the cobordism conjecture \cite{McNamara:2019rup}, ETW branes always exist and bubbles of nothing should hence be considered as a potential decay process for all string models with broken SUSY \cite{GarciaEtxebarria:2020xsr}.

Specifically for type IIB CY orientifold compactifications with O3/O7 planes, which are most relevant in the present context, an ETW brane was constructed in \cite{Friedrich:2023tid}. It is based on an O5 orientifolding reflecting one of the non-compact dimensions. The preliminary analysis of the tension suggests that the induced decays to nothing occur faster than tunneling to decompactification and generic flux transitions, but are nevertheless unlikely to compete with the destruction of the uplift. It would, however, be important to understand whether other ETW branes with more negative tension and hence higher decay rates exist.

\section{Explicit predictions in the (string theory) multiverse}\label{sect_explicit_predictions}
We now return to the study of anthropic predictions in the multiverse. While much of what follows is generic and applies to any underlying landscape, some of the arguments used in this section rely on the string-specific results of Sect.~\ref{sect_String_rates}.
As discussed in Sect.~\ref{sect_anthropic_pred_WDW}, $p({\rm obs},i)$, the probability of finding an observer in vacuum $i$, is mainly determined by the value $p_{{\rm inf }(i)}$. This is the quantity we will now try to calculate on the basis of \eqref{p_i_solution}.
\subsection{Creation from nothing or tunneling from a higher-energy dS?}\label{sect_direct_creation_or_tunneling}

In the solution to the rate equation \eqref{p_i_solution}, the term $p_{i,N}$ corresponds to the contribution where $i$ is reached via $N$ tunneling transitions after creation from nothing. 
Each transition comes with an additional suppression which may be approximately quantified by a factor
\begin{align}
    \frac{\sum_{k\in {\rm dS}}\Gamma_{j\to k}}{\Gamma_j}=\frac{\sum_{k\in {\rm dS}}\Gamma_{j\to k}}{\sum_{k\in {\rm dS}} \Gamma_{j\to k}+ \sum_{y\in {\rm Terminals}} \Gamma_{j\to y}}\,.\label{branching_sum}
\end{align}
Here, $j$ denotes a fixed dS vacuum, $k$ runs over all other dS vacua and $y$ runs over all terminal vacua and `nothing'.
In the KKLT/LVS landscape analyzed in Sect.~\ref{sect_String_rates}, dS-dS transitions are generally maximally suppressed. By contrast, transitions to decompactification, to nothing, and to AdS are much faster. 
The fact that dS-dS transition rates are strongly suppressed compared to the total dS decay rate could be a general feature of the string landscape and it would be interesting to explore this further.
In addition, from the difficulties of constructing dS vacua in string theory, one may expect that they are vastly outnumbered by AdS vacua. We conclude that $\sum_{k\in {\rm dS}}\Gamma_{j\to k}/\Gamma_j\ll 1$ holds for almost all dS vacua $j$.\footnote{A minor exception arises because we treat inflationary plateaus as separate dS vacua. If $j$ stands for such a plateau, the factor \eqref{branching_sum} is obviously very close to unity in cases where the rolling takes us to a metastable dS minimum. Since inflationary plateaus are presumably rare, we consider it justified to ignore this subtlety.
}
As a result, the sum \eqref{p_i_solution} converges quickly and we may use the approximation
\begin{align}
    p_i\simeq p_{i,0}+p_{i,1}\,.\label{p_i_p0_p1}
\end{align}
In this approximation, vacuum $i$ can either be directly created from nothing (corresponding to the term $p_{i,0}$) or via one intermediate tunneling process (corresponding to $p_{i,1}$).

We now pick an anthropically viable vacuum $i$ and apply \eqref{p_i_p0_p1} to its inflationary plateau which leads to:
\begin{align}
    p_{{\rm inf}(i)}\,\,\simeq\,\, p_{{\rm inf}(i),\,0}\,+\,p_{{\rm inf}(i),\,1}\,\,=\frac{1}{\Gamma_{{\rm inf}(i)}}\left(\,\,
	J_{{\rm inf}(i)} \,\,+\,\, \sum_{o\,\neq\,{{\rm inf}(i)}} J_o\frac{\Gamma_{o\to {\rm inf}(i)}}{\Gamma_o}\right)\,.
    \label{two_creation_ways}
\end{align}
The relative size of the two terms in parenthesis
determines whether the direct creation from nothing or tunneling in the background of another dS space is the most likely process to induce reheating in vacuum $i$.
The summation index $o$ in the second term indicates that we sum over all vacua `other' than inf($i$). 
From \eqref{p_i_solution}, we observe that each $p_i$ contains an overall factor $1/\Gamma_i$. The appearance of this factor is intuitively clear: A smaller decay rate makes a vacuum longer-lived and thus increases the probability of {\it finding} it in the multiverse. However, as we discussed in Footnote \ref{footnote_p_obs}, $p({\rm obs},i)$ should be proportional to the probability of {\it starting} inflation in inf($i$) (leading to reheating in vacuum $i$) and should hence be independent of $\Gamma_{{\rm inf}(i)}$. An improved estimate of $p({\rm obs},i)/p({\rm obs},j)$, with $i,j$ anthropically viable vacua, is thus given by 
\begin{align}
    \frac{p({\rm obs},i)}{p({\rm obs},j)}\simeq \frac{\Gamma_{{\rm inf}(i)}\,p_{{\rm inf}(i)}}{\Gamma_{{\rm inf}(j)}\,p_{{\rm inf}(j)}}\,.\label{p_obs_refined}
\end{align}
Although we expect $\Gamma_{{\rm inf}(i)}/\Gamma_{{\rm inf}(j)}$ to be a subleading effect and thus \eqref{p_obs_refined} to be only a minor refinement of \eqref{anthr_prediction}, we shall employ the refined ratio of probabilities \eqref{p_obs_refined} from here on.

In what follows, it will be convenient to work mostly with the decay exponents $B$ rather than with the decay rates $\Gamma$. We will label these exponents in the same way as rates, i.e.
\begin{equation}
\Gamma_i \propto \exp(-B_i)\,,\qquad
\Gamma_o \propto \exp(-B_o)\,,\qquad \Gamma_{o\to{\rm inf}(i)}\propto
\exp(-B_{o\to{\rm inf}(i)})\,,\qquad \mbox{etc.}
\end{equation}
As before, we denote by $S_i$ the entropy of vacuum $i$ and by $\mathcal{S}_i$ the euclidean action of the instanton responsible for the creation of vacuum $i$ from nothing.
In this notation, the inequality
\begin{align}    
    0<B_o < B_{o\to {{\rm inf}(i)}}<S_o\label{estimates}
\end{align}
always holds.
From the analysis of Sect.~\ref{sect_String_rates}, we learned that in a KKLT/LVS landscape, generic transitions between dS vacua are close to being maximally suppressed while decays to AdS can proceed via the destruction of the uplift and are hence much faster. In this type of landscape, we have the refined statement
\begin{align}
    0<B_o\ll B_{o\to {{\rm inf}(i)}}\simeq S_o\,.\label{estimates_generic}
\end{align}

Evidently, in a landscape without eternal inflation, tunneling is a subdominant effect such that inflationary vacua are mainly created directly from nothing with rate $J_{{\rm inf}(i)}$.

\subsubsection{Hartle-Hawking-type creation}
When adopting the sign choice of Hartle and Hawking, i.e.~$J\sim \exp(-\mathcal{S})$, Bubble-of-Something and Boundary creation processes are always subdominant and we can focus on the standard creation of a spherical universe from nothing. Since the terms in the sum of $p_{\rm inf(i),1}$ \eqref{two_creation_ways} are all exponentially large or small, we approximate the sum by its largest term. We then have
\begin{align}
    \frac{p_{\rm inf(i),1}}{p_{\rm inf(i),0}}=\frac{J_o\,\Gamma_{o\to {{\rm inf}(i)}}/\Gamma_o}{J_{{\rm inf}(i)}}\simeq \exp(S_o-B_{o\to {{\rm inf}(i)}}+B_o-S_{{\rm inf}(i)})\,.
    \label{p10hh}
\end{align}
If $S_{{\rm inf}(i)}>S_o$, it is immediately clear that the direct creation of the inflationary state inf$(i)$ from nothing is more likely than tunneling to that state from vacuum $o$.

If $S_{{\rm inf}(i)}<S_o$, i.e.~when uptunneling needs to occur, more thought is required:  In this case, we may use the general identity\footnote{
This 
can fail if, as argued in \cite{Farhi:1989yr,Blanco-Pillado:2019xny} (see also \cite{DeAlwis:2019rxg,Cespedes:2020xpn}), detailed balance does not hold such that up-tunneling rates are much larger than naively expected. The relevant transitions are such that, from the outside observer's perspective, the created vacuum is behind a black hole horizon. It is not clear how to treat this in the local WDW framework and we leave this issue to future work.
}
$B_{o\to {{\rm inf}(i)}}=S_o-S_{{\rm inf}(i)}+B_{{{\rm inf}(i)}\to o}$ to simplify \eqref{p10hh} as follows:
\begin{align}
    \frac{p_{\rm inf(i),1}}{p_{\rm inf(i),0}}= \frac{J_o\,\Gamma_{o\to {{\rm inf}(i)}}/\Gamma_o}{J_{{\rm inf}(i)}}\simeq \exp(-B_{{{\rm inf}(i)}\to o}+B_o)\,.\label{p10hh_up}
\end{align}
As shown in Sect.~\ref{sect_String_rates}, the KKLT/LVS landscape generically has $B_{\rm inf(i)\to o}\simeq S_{{\rm inf}(i)}$. By contrast, $B_o$ is in the field-theoretic regime, determined by the destruction of the uplift (see Sect.~\ref{ssec:uplift}).
As discussed below \eqref{B_KPV_main}, especially when considering the $\overline{D3}$-brane uplift, $B_o$ cannot be very large. Hence, one generally expects $B_o< B_{{{\rm inf}(i)}\to o}$. Let us be more specific about this point:

CMB data restricts the tensor to scalar ratio to $r<0.04$ \cite{Planck:2018jri} leading to a bound on the Hubble scale during the inflationary era that occurred in our own past: $H_{\rm inflation}\lesssim 0.4 \cdot 10^{-4}M_P$. Hence the inflationary plateau of our universe satisfies $S_{\rm inflation}=8\pi^2M_P^2/H_{\rm inflation}^2\gtrsim 5\cdot 10^{10}$.
For consistency of the uplift (see e.g.~\cite{Carta:2019rhx,Gao:2020xqh,McAllister:2024lnt,Junghans:2022exo, Gao:2022fdi, Junghans:2022kxg, Hebecker:2022zme, Schreyer:2022len, Schreyer:2024pml}), the flux number $M$ in \eqref{B_KPV_main} cannot be larger than $\order{100}$.
As a result,\footnote{
Note that the warp factor suppressing the uplift is $\exp(8\pi N/3g_sM^2)$. If we very generously require this to only be $\exp(-10)$ and restrict the tadpole by $N<10^3$, we find $g_sM^2\lesssim 10^3$. Then \eqref{B_KPV_main} gives $B_{\rm KPV}\lesssim 3\times 10^8$.
}
$B_{\rm KPV}\lesssim 10^8$.
Recalling the great difficulties in realizing a metastable uplift (briefly mentioned below \eqref{B_KPV_main}), it becomes clear that our assumptions are very generous and that, for inflationary plateaus that are observationally viable, $B_o<B_{{{\rm inf}(i)}\to o}\sim S_{{\rm inf}(i)}$ appears unavoidable. 
As a result, direct creation from nothing is the dominant way of creating inflationary universes of such type.

However, for inflationary plateaus at very high energies and for vacua $o$ with a relatively stable uplift (possibly not the $\overline{D3}$-uplift), $B_o> B_{{{\rm inf}(i)}\to o}$ appears to be generally possible.
Moreover, if non-generic flux transitions between dS vacua are relevant, $B_{{\rm inf}(i)\to o}< S_{{\rm inf}(i)}$ and thus $B_o>B_{{\rm inf}(i)}\to o$ can also be achieved. Finally, we could be dealing with a landscape which is very different from any KKLT/LVS-based expectations. Also in this case $B_o>B_{\rm inf(i)\to o}$ could hold for certain inflationary plateaus and other vacua $o$. In these cases, the inflationary plateau ${\rm inf}(i)$ is more likely to arise through tunneling than through direct creation from nothing \`{a} la Hartle and Hawking.

In summary, expectations based on an LVS/KKLT-type landscape imply that for the Hartle-Hawking sign choice inflationary plateaus are directly created from nothing rather than through tunneling events. A key possibility for how this might fail is if low-scale dS vacua are much more long-lived than the known fragile uplifts suggest.

\subsubsection{Linde/Vilenkin creation with nontrivial topology}
In case the creation of toric or other universes with nontrivial topology is considered, $J_o\simeq J_{{\rm inf}(i)}$ holds to leading order for all $o$. As a result, the direct creation from nothing of vacuum $i$ in an inflationary state is the most likely scenario (recall, however, the caveat mentioned at the end of Sect.~\ref{topsec}).

\subsubsection{Linde/Vilenkin creation with spherical topology}
In this approach, the Euclidean action $\mathcal{S}$ of the creation instanton is related to the creation rate by $J\sim\exp(+\mathcal{S})$. If no ETW branes exist, we have $-\mathcal S = |\mathcal S|=S_{dS}$. Otherwise, a bubble of something or a boundary instanton is the most likely creation process, with $- \mathcal S = |\mathcal S|<S_{dS}$ given in \eqref{S_bos},\eqref{S_boundary}. Let us start with the part of analysis which applies to both cases -- with and without ETW branes:

As before, a key quantity of interest is the ratio
\begin{align} \label{ratio_LV}
    \frac{p_{{\rm inf}(i),1}}{p_{{\rm inf}(i),0}}=\frac{J_o\,\Gamma_{o\to {{\rm inf}(i)}}/\Gamma_o}{J_{{\rm inf}(i)}}\simeq \exp(-|\mathcal{S}_o|-B_{o\to {{\rm inf}(i)}}+B_o+|\mathcal{S}_{{\rm inf}(i)}|)\,.
\end{align}
Let us first see under which conditions, for given ${\rm inf}(i)$, there exists a vacuum $o$ making this ratio large. Clearly, we have to focus on vacua $o$ with small $|\mathcal{S}_o|$. We may then assume $|\mathcal S_o|<|\mathcal S_{{\rm inf}(i)}|$.\footnote{This of course requires that, for all anthropically viable vacua $i$, vacua $o$ satisfying $|\mathcal{S}_o|<|\mathcal{S}_{{\rm inf}(i)}|$ exist. We believe this is a good assumption since most vacua are likely not inflationary plateaus leading to an anthropically viable vacuum.}
Using \eqref{estimates}, we can estimate
\begin{align}
    -|\mathcal S_o|-B_{o\to {{\rm inf}(i)}}+B_o+|\mathcal S_{{\rm inf}(i)}| > -|\mathcal S_o|-B_{o\to {{\rm inf}(i)}}+|\mathcal S_{{\rm inf}(i)}| > -(|\mathcal S_o|+S_o)+|\mathcal S_{{\rm inf}(i)}|\,.\label{creation_estimate}
\end{align}
We conclude that the condition
\begin{align}
    |\mathcal{S}_{{\rm inf}(i)}|>S_o+|\mathcal{S}_o|\label{S_ineq1}
\end{align}
is sufficient to ensure that $p_{{\rm inf}(i),1}/p_{{\rm inf}(i),0}\gg 1$ and hence that ${\rm inf}(i)$ is more likely to arise through tunneling from $o$ than via creation from nothing.

Let us now ask under which conditions $p_{{\rm inf}(i),1}/p_{{\rm inf}(i),0}$ always remains small. For this, we recall that in the KKLT/LVS landscape \eqref{estimates_generic} holds. This implies
\begin{align}
    -|\mathcal S_o|-B_{o\to {{\rm inf}(i)}}+B_o+|\mathcal S_{{\rm inf}(i)}| \simeq -|\mathcal S_o|-S_o+|\mathcal S_{{\rm inf}(i)}|\,.
\end{align}
We conclude that
$p_{{\rm inf}(i),1}/p_{{\rm inf}(i),0}\ll 1$ and hence that that inf($i$) is most likely to be created directly from nothing if
\begin{align}
	|\mathcal{S}_{{\rm inf}(i)}|< S_o+|\mathcal{S}_o|\label{S_ineq2}
\end{align}
holds for all $o$.

To be more specific, we have to distinguish different cases:
\paragraph{No (relevant) ETW branes:}
As discussed around Eq.~\eqref{S_bos},\eqref{S_boundary}, if there are no branes satisfying $T_4\ell_{{\rm inf}(i)}\gg M_P^2$ or $0<-T_4\ell_{{\rm inf}(i)}\ll M_P^2$ (or if ETW branes do not exist at all), then $|\mathcal{S}_{{\rm inf}(i)}|$ is of order $S_{{\rm inf}(i)}$. Moreover, $S_{{\rm inf}(i)}$ is expected to be much larger than the smallest $S_o$ in the landscape. Now, using the fact that $0<|\mathcal{S}_o|<S_o$, we arrive at the conclusion that \eqref{S_ineq1} always holds and thus the creation of inf($i$) in the background of some $o$ is more likely than the direct creation of inf($i$) from nothing.
\paragraph{Relevant ETW branes:} 
If ETW branes with $T_4\gg M_P^2/\ell_{{\rm inf}(i)}$ exist, $|\mathcal{S}_{{\rm inf}(i)}|$ is given by the bubble of something result \eqref{large_T}
\begin{align}
|\mathcal{S}_{{\rm inf}(i)}|\simeq 8\pi^2\frac{M_P^6}{T_4^2}\,.
\end{align}
It is then possible that \eqref{S_ineq2} holds for all vacua $o$, making the direct creation from nothing most likely. More concretely, if the landscape contains a vacuum with minimal de Sitter radius $\ell_{min}$ (i.e.~with the largest cosmological constant), then condition \eqref{S_ineq2} is always fulfilled if $T_4\gg M_P^2/\ell_{min}$.
Clearly, an ETW brane tension is similar to a 3d cosmological constant. 
From what we know about the landscape, one may argue that it is easier to obtain large energy densities in 3d than in 4d. Hence it is quite plausible that \eqref{S_ineq2} does indeed hold for many inflationary plateaus inf$(i)$, which are hence predominantly created by a bubble-of-something process. Obviously, this holds true if the landscape has no dS vacua.

Alternatively, if ETW branes with $0<-T_4\ell_{{\rm inf}(i)}\ll M_P^2$ exist, the boundary proposal becomes relevant and the corresponding euclidean action is given by \eqref{S_bdry_small_T}:
\begin{align}
    |\mathcal{S}_{{\rm inf}(i)}|\simeq 4\pi^2|T_4|\ell_{{\rm inf}(i)}^3\,.
\end{align}
Again, the direct creation from nothing can easily be the dominant process leading to a cosmology which starts on the inflationary plateau inf$(i)$.  In particular, this will be the case if the landscape has no dS vacua and if even positive-tension ETW branes are forbidden by Swampland constraints. 

We see that ETW branes can drastically change phenomenological predictions in the multiverse. This is particularly true in the presence of strong Swampland constraints on models with positive energy density.

\subsection{Most likely scale of inflation}
Let us now try to use \eqref{anthr_prediction} and the dependencies of creation and decay rates on the vacuum energies to make an anthropic prediction for the scale of inflation.\footnote{
Our work is focused on the energy scale at the beginning of inflation. We shall assume that it is similar to the phenomenologically relevant scale associated with the CMB. Clearly, in principle there are models where these two scales are very different.}
\subsubsection{Hartle-Hawking-type creation}
We have discussed below \eqref{p10hh} that most inflationary plateaus, and especially the ones that are still in the observationally viable regime, get dominantly created directly from nothing.
Using \eqref{p_obs_refined} and \eqref{p_i_solution} to zeroth order, we find
\begin{align}
   	p({\rm obs},i)\propto J_{{\rm inf}(i)}\propto \exp(S_{{\rm inf}(i)})=\exp(24\pi^2M_P^4\,/\,V_{{\rm inf}(i)})\,.\label{Predict_Ha_Ha}
\end{align}
We see that the landscape dynamics gives an exponential bias towards low-scale inflation.
The details of tunneling transitions do not play an important role.

This result leads to an immediate tension with observations, as has recently been made explicit in \cite{Maldacena:2024uhs}. The exponential preference for a low scale of inflation leads to an overwhelming probability for having the shortest possible period of inflation that is still compatible with the formation of anthropic observers. However, CMB measurements such as the observed spatial flatness of our Universe show inflation lasted longer than anthropically required. Therefore, creation out of nothing in the Hartle-Hawking scenario appears to be incompatible with data.

In the discussion below \eqref{p10hh}, we have specified what precisely it would take to find inflationary plateaus which are populated through tunneling: At least in the KKLT/LVS landscape, they can only exist at very high energies or if very stable uplifts can be realized. Let us nevertheless try to understand whether such plateaus (if existent) can be anthropically preferred. For this purpose, we assume that inf($i$) is a low-energy plateau with dominant creation directly from nothing and inf($j$) a high-energy plateau with dominant creation in the background of some high entropy dS $o$. To compare them, we use
\eqref{p_obs_refined},\eqref{p_i_solution} and obtain
\begin{align}
    \frac{p({\rm obs},i)}{p({\rm obs},j)}\simeq \frac{\Gamma_{{\rm inf}(i)} p_{\rm inf(i),0}}{\Gamma_{{\rm inf}(j)} p_{{\rm inf}(j),1}}=\frac{J_{{\rm inf}(i)}}{J_o \Gamma_{o\to {{\rm inf}(j)}}/\Gamma_o}\simeq \exp(S_{{\rm inf}(i)}-S_o+B_{o\to {{\rm inf}(j)}}-B_o)\,.
\end{align}
Employing the relation $B_{o\to {{\rm inf}(j)}}=S_o-S_{{\rm inf}(j)}+B_{{{\rm inf}(j)}\to o}$ this simplifies to
\begin{align}
     \frac{p({\rm obs},i)}{p({\rm obs},j)}\simeq \exp(S_{{\rm inf}(i)}-S_{{\rm inf }(j)}+B_{{\rm inf}(j)\to o}-B_o)\,.
\end{align}
At least in the KKLT/LVS landscape, we expect $p({\rm obs},i)/p({\rm obs},j)\gg 1$ since $S_{{\rm inf}(i)}\gg S_{{\rm inf}(j)}+B_o$ is expected for low-energy plateaus inf($i$) (recall the discussion below \eqref{p10hh_up}).
As a result, low-scale inflationary regions directly created from nothing are most likely to be realized and the conclusions below \eqref{Predict_Ha_Ha} still hold.\footnote{These conclusions, including the claim that the Hartle-Hawking sign choice is ruled out by observations, could be altered by detailed-balance-violating, fast upward transitions such as \cite{Farhi:1989yr,Blanco-Pillado:2019xny}. It would be interesting to study how such transition should be treated in the local WDW framework.}

\subsubsection{Linde/Vilenkin creation with nontrivial topology:}\label{sect_scale_of_inflation_tori}
If the creation of small universes with non-trivial topology is possible, then all (quasi-)dS vacua are equally likely to be created from nothing. This expectation holds at leading order, i.e.~disregarding non-exponential effects, and in the 4d EFT. Accepting these caveats, one may conclude that the distribution of the scale of inflation measured by anthropic observers is equal to the distribution of the number of inflationary plateaus in the landscape as a function of energy. Such a distribution has been estimated for a toy-model of the string landscape \cite{Pedro:2013nda}.\footnote{Note that~\cite{Pedro:2016sli} proposed to require both sufficient slow-roll inflation \emph{and} rolling into an anthropically viable state thereafter. The number frequency distribution of inflation together with such a `graceful exit' can differ strongly from what one obtains by just asking for sufficient slow-roll inflation.}
It was found that the number frequency is highest for small-field models of plateau inflation, which are favored by recent cosmological measurements \cite{Planck:2018jri, BICEP:2021xfz}. 
Among the small-field scenarios, those with the largest scale of inflation appear to be preferred \cite{Pedro:2013nda}.

Key challenges for the future are the analysis of the creation with non-trivial topology at sub-exponential level and beyond the 4d EFT, i.e.~including effects of the compact space. Moreover, we obviously need the distribution of inflationary plateaus in the actual string landsape.

\subsubsection{Linde/Vilenkin type creation with spherical topology:}
For the Linde/Vilenkin type creation with spherical topology, we again distinguish between the following cases:
\paragraph{No (relevant) ETW branes:}
If there are no relevant ETW branes to significantly change the traditional Linde/Vilenkin result, the creation of $i$ in the background of some other vacuum $o$ is the most likely scenario. 
Hence, observers in the multiverse could measure signatures of bubble collision events \cite{Chang:2008gj,Kleban:2011pg}.

Quantitatively, we find at leading order
\begin{align}
p({\rm obs},i)
\,\propto\,
	\Gamma_{{\rm inf}(i)}p_{{\rm inf}(i)} 
 \,\simeq\, J_o\frac{\Gamma_{o\to {\rm inf}(i)}}{\Gamma_o}\,\simeq\, \exp(-\frac{24\pi^2 M_P^4}{V_o})\frac{\Gamma_{o\to {\rm inf}(i)}}{\Gamma_o}\,.\label{Predict_LV}
\end{align}
Clearly, it is favorable to nucleate a vacuum with large energy density from nothing. 
Since $\Gamma_o$ is dominated by the destruction of the uplift, which happens in the field-theoretic regime, it is independent of the vacuum energy.
The tunneling rate $\Gamma_{o\to {\rm inf}(i)}$ becomes a key player in making anthropic predictions.
This is in contrast to \eqref{Predict_Ha_Ha} where solely the energy of the inflationary plateau determines the prediction.
Eq.~\eqref{Predict_LV} favors those inflationary plateaus which are most closely connected to the highest-energy vacua of the landscape.
In the KKLT/LVS landscape of Sect.~\ref{sect_String_rates}, the generic dS-dS tunneling rate is close to being maximally suppressed, cf.~\eqref{B_flux}.
As a result, we do not expect a strong correlation between the energy densities ${\rm inf}(i)$ and $o$ in the string landscape. 
Therefore, the prediction for the scale of inflation may turn out to be similar to that of Sect.~\ref{sect_scale_of_inflation_tori}, as a result of the creation of universes with non-trivial topology.
However, this may change if there are many dS-dS transitions which are not maximally suppressed, such that the energy dependence of the tunneling rate $\Gamma_{o\to {\rm inf}(i)}$ in \eqref{Predict_LV} becomes important. For example, as discussed in Sect.~\ref{ssec:flux_transition}, non-generic flux transitions may lead to such fast dS-dS transitions. Moreover, 
dS-dS tunneling rates could naturally be important if the relevant landscape is very different from expectations based on the flux-landscape of type-IIB supergravity. 

\paragraph{Relevant ETW branes:} 
If bubble of something creation events are dominant, the prediction becomes
\begin{align}
p({\rm obs},i)
\,\simeq\,
	\Gamma_{{\rm inf}(i)}p_{{\rm inf}(i)} 
 \,\simeq\, J_{{\rm inf}(i)}\,\simeq\, \exp(-\frac{8\pi^2M_P^4}{T_4^2})\,.
\end{align}
In this case, a new quantity comes in to determine the most probable scale of inflation: It is neither the energy density of the plateau nor the tunneling rates towards ${{\rm inf}(i)}$, but rather it is the tension of the ETW brane which may exist in the quantum-gravity EFT describing the  inflationary plateau.

If the boundary proposal is relevant, a similar situation occurs:
\begin{align}
p({\rm obs},i)
\,\simeq\,
	\Gamma_{{\rm inf}(i)}p_{{\rm inf}(i)}  
 \,\simeq\, J_{{\rm inf}(i)}\,\simeq\,
        \exp(-4\pi^2|T_4|\ell_{{\rm inf}(i)}^3)\,.
\end{align}
Now those inflationary plateaus are favored which allow for an ETW brane with negative tension and with the smallest values of $|T_4|\ell_{{\rm inf}(i)}^3$.

The above results are novel predictions in the multiverse clearly showing that relations between ETW branes and cosmology deserve further studies.
It would be especially important to uncover any possible correlations between the values of $T_4$ and features of the corresponding inflationary plateau, such as its energy scale.
\paragraph{No eternal inflation:} If quantum gravity does not allow for eternal inflation, direct creation of an inflationary state from nothing is obviously preferred. If the traditional Linde/Vilenkin proposal is relevant, this favors the largest anthropically viable inflationary scale. 
If creation proposals including ETW branes are relevant, the same conclusions hold as described above.

\section{Conclusions}
A measure is needed both for `rocky' landscapes, with many dS vacua and eternal inflation, as well as for `swampy' landscapes, with cosmological expansion mostly or exclusively driven by some form of decaying dark energy. For both scenarios, we have developed and applied the `local WDW measure', resting on two ingredients: First, the assumption that (quasi-)dS vacua have finite-dimensional Hilbert spaces (the Cosmological Central Dogma or CCD). Second, the expectation that in this context the wave function of the universe $\Psi$ follows from solving an inhomogeneous version of the Wheeler-DeWitt (WDW) equation, which accounts for creation from nothing.
Predictions are made by projecting $\Psi$ on the subspace of interest.

Concretely, we focused on anthropic observers living near post-inflationary reheating surfaces and developed the local WDW measure to the point where explicit predictions for the scale of inflation become possible. It would be interesting to similarly employ the local WDW measure for the study of alternative cosmological approaches, such as \cite{Khoury:2001wf, Ijjas:2018qbo}.
An obstacle for making predictions is the fact that, in the local WDW framework,  $\Psi$ is not normalizable.
Hence, predictions of relative probabilities like 
$\bra{\Psi} P_{\alpha} \ket{\Psi} / \bra{\Psi}P_{\beta} \ket{\Psi}$, with $P_{\alpha,\beta}$ appropriate projectors, may in general be ill-defined. 
We argued that anthropic predictions can nevertheless be possible because $\Psi$ is always projected on the subspace $\mathcal{H}_{obs}$ in which observers are present. We then investigated whether this projected state is normalizable.

Clearly, if the landscape has finitely many (quasi-)dS vacua, which by assumption all have finite-dimensional Hilbert spaces, then the projected wave functions arising in this context are normalizable. While, by contrast, AdS has an infinite-dimensional Hilbert space, cosmological AdS always crunches. We argue that in the cosmological transition from inflation to AdS and to the crunch only finitely many independent observers can appear. The subspace of ${\cal H}_{obs}$ associated with observers in AdS is then nevertheless finite-dimensional. The most problematic case is that of Minkowski space: In a dS-Minkowski transition, an infinite reheating surface with infinitely many observers can form which, moreover, can all be seen from the infinite future. 
We argued that the CCD nevertheless suggests that only a finite portion of this surface and hence a finite number of the observers can be considered as independent.
We provided a set of independent arguments supporting this claim.
In addition, we also analyzed how predictions in the multiverse change if one considers the infinitely many observers in Minkowski space as independent.
In particular, this leads to the prediction that our observed dark energy should decay. 
All these arguments are summarized in Sect.~\ref{sect_summary_on_counting}. To clarify whether our suggested counting of observers is correct, the CCD and its application to quasi-dS, i.e.~inflationary universes, needs to be studied in more detail.
A key problem appears to be the interplay with the infinite-dimensional Hilbert space of asymptotic Minkowski cosmologies.

Assuming the projection of $\Psi$ on $\mathcal{H}_{obs}$ to be normalizable, we then proceeded towards our goal of predicting the scale of inflation. The result turned out to significantly depend on the creation rates of universes from nothing. Such an `initial conditions dependence' is expected for local measures.  In `swampy' landscapes, creation from nothing is of overwhelming importance and tunneling transitions are subdominant.
For `rocky' landscapes, creation and tunneling rates together determine the predictions.

The creation rates depend strongly on the choice of sign in the relation between instanton action and creation amplitude. When adopting the `Hartle-Hawking' choice, one finds that our universe was created directly from nothing.
Moreover, the scale of inflation is predicted to be as low as anthropically possible, in strong tension with observation.

When adopting the `Linde/Vilenkin' choice, different vacuum creation mechanisms need to be considered. For example, it is possible that small compact universes with zero or negative curvature (e.g.~3-tori) are as likely to fluctuate out of nothing as small 3-spheres.
It is then a prediction that we live in one of these spatially compact universes with non-trivial topology. Since there are no obvious exponential suppression effects, the scale of inflation is predicted on the basis of vacuum counting, i.e., it is in the energy-region where most inflationary plateaus of the landscape can be found.
While it would be important to study creation with non-trivial topology in more detail, this is challenging since one expects such processes to be UV-sensitive.
If the creation of universes with non-trivial topology is suppressed and no relevant ETW branes exist, we predict that our inflationary plateau was populated through a tunneling process from a high-scale dS vacuum. 
The most likely scale of inflation is again determined by the number density of inflationary plateaus in the landscape.
If, by contrast, ETW branes with appropriate tensions exist, we predict that a universe on our inflationary plateau has been created directly from nothing. 
The scale of inflation is then determined by correlations between this energy scale and the availability and tension of ETW brane involved in this (`bubble of something' or `boundary') creation process. ETW branes may hence drastically affect predictions in cosmology. Correlations between their tension and observable features of our (inflationary) vacuum represent a key research target.

Let us finally state some explicit but preliminary connections to observations:
The Hartle-Hawking sign choice for creation from nothing appears to be observationally strongly disfavored. 
However, this could change 
if fast, detailed-balance-violating 
up-tunneling transitions can be established.
When using the sign choice of Linde and Vilenkin, both direct creation from nothing and tunneling dynamics can be important.
Nevertheless, in many cases anthropic predictions can be dominated simply by the available number of vacua of a certain type in the landscape. This includes the particularly important case in which small universes with non-trivial topology can be created from nothing.
A detailed study of the distribution of the scale of inflation in the landscape would hence be important (for first steps see \cite{Pedro:2013nda}).

The tensor-to-scalar ratio $r$, related to the amplitude of primordial gravitational waves, is an observable directly linked to the energy scale of inflation (see e.g. \cite{CMB-S4:2016ple}):
\begin{equation}
    V_{{\rm inf}(i)}^{1/4} \approx 10^{16}\, \text{GeV} \left(\frac{r}{0.01}\right)^\frac{1}{4} \,.
\end{equation}
So far, no detection has been made and the current bound \cite{Planck:2018jri, BICEP:2021xfz} $r\lesssim 0.04$ restricts the scale of inflation to $V_{{\rm inf}(i)}^{1/4}\lesssim 1.4\cdot 10^{16}$GeV.
One may then conclude that, while the distribution of inflationary plateaus in the landscape cannot be peaked at energy scales above $10^{16}$GeV, a peak at any lower energy value is consistent with present data. This matches with the results of \cite{Pedro:2013nda}, where 
a toy-model for the plateau distributions in the type-IIB
flux landscape was studied, finding a preference for
$r\sim 10^{-3}$ (detectable with planned CMB measurements such as CMB-S4 \cite{Abazajian:2019eic} and LiteBIRD \cite{LiteBIRD:2022cnt}).

\subsection*{Acknowledgements}
We thank Sebastian Zell for being part of the collaboration during the early stages of this project and for significantly contributing to the development of this work. 
We are furthermore grateful for useful discussions with Raphael Bousso and Thibaut Coudarchet. This work was supported by Deutsche Forschungsgemeinschaft (DFG, German Research Foundation) under Germany’s Excellence Strategy EXC 2181/1 - 390900948 (the Heidelberg STRUCTURES Excellence Cluster). AW is partially supported by the Deutsche Forschungsgemeinschaft under Germany’s Excellence Strategy - EXC 2121 “Quantum Universe” - 390833306, by the Deutsche Forschungsgemeinschaft through a German-Israeli Project Cooperation (DIP) grant “Holography and the Swampland”, and by the Deutsche Forschungsgemeinschaft through the Collaborative Research Center SFB1624 ``Higher Structures, Moduli Spaces, and Integrability’'. 
\appendix

\section{Solving the rate equation}\label{appendix_solving_rate_equation}
Solving the rate equation \eqref{rate_equation} amounts to inverting the matrix $M=X+Y$ with
\begin{align}
    X_{ij}=\delta_{ij}\Gamma_i\,,\quad Y_{ij}=-\Gamma_{j\to i}\,.
\end{align}
Formally, the inverse is given by
\begin{align}
	(X+Y)^{-1}=X^{-1}-X^{-1}YX^{-1}+X^{-1}YX^{-1}YX^{-1}-...\,.\label{lemma_inverse_of_matrix_sum}
\end{align}
To see that this series converges and $M$ is hence indeed invertible (an observation also made in 
\cite{Garriga:2005av}), we write
\begin{align}
    (YX^{-1})_{ij}=-\sum_k \Gamma_{k\to i}\frac{\delta_{kj}}{\Gamma_j}=-\frac{\Gamma_{j\to i}}{\Gamma_j}\,,\label{calculation_B_Ainverse}
\end{align}
and observe that
\begin{align}
	\forall i,j: \qquad \sum_i|(YX^{-1})_{ij}|<1\,.
\end{align}
This crucial inequality is a consequence of the sum going only over (approximate) dS solutions, excluding terminal vacua.

Given that the standard matrix norm $\Vert A\Vert\equiv \sup_x (\Vert Ax\Vert/\Vert x\Vert )$ satisfies 
$\Vert A\Vert = \max_j\sum_i|A_{ij}|$,
we may now conclude that $\Vert YX^{-1}\Vert<1$. Hence the 
series  \eqref{lemma_inverse_of_matrix_sum} converges and \eqref{p_i_solution} solves our rate equation.

\section{The decay rate in certain limits}\label{Appendix_rate_discussion}
For later use, we collect some facts about the tunneling rate \eqref{decay_rate} and consider certain limits.
First, we note that $V_f$ and $V_t$ can be expressed using the dimensionless variables \eqref{dimensionless_variables} as
\begin{align}
    V_f=\frac{3T^2(y+1)}{8M_P^2x}\,,\qquad V_t=\frac{3T^2(y-1)}{8M_P^2x}\,.
\end{align}
The entropy of the false vacuum may be written as
\begin{align}
    S_f=\frac{64\pi^2 M_P^6x}{T^2(y+1)}\,.\label{appendix_entropy_dimless_variables}
\end{align}
Using
\begin{align}
    \frac{27 T^6}{(V_f-V_t)^3}=64M_P^6x^3
\end{align}
and \eqref{appendix_entropy_dimless_variables}, the decay exponent \eqref{decay_rate} can be given in the form
\begin{align}
    B=\frac{64M_P^6x^3}{2T^2}r(x,y)=S_f\frac{x^2 (y+1)}{2}r(x,y)=S_f\frac{1+xy-\sqrt{1+2xy+x^2}}{(y-1)\sqrt{1+2xy+x^2}}\,.\label{appendix_general_rate}
\end{align}
\paragraph{Tunneling to Minkowski:}
The tunneling process to Minkowski space is of interest for example in the decay to decompactification. $V_t=0$ is equivalent to $y=1$ for which $B$ becomes
\begin{align}
    B=S_f\frac{x^2}{(1+x)^2}=\frac{S_f}{(1+1/x)^2}=\frac{S_f}{\left(1+\frac{4M_P^2V_f}{3T^2}\right)^2}\,.\label{appendix_Minkowski_tunneling}
\end{align}
We see that for large $x$, $B\approx S_f$ and the lifetime of the dS space approaches its recurrence time.

\paragraph{Large $x$ expansion:}
For $x\gg y$ and $x\gg 1$, we may characterize $B$ by expanding  the function of $x$ and $y$ appearing in \eqref{appendix_general_rate} in $1/x$:
\begin{align}
    B=S_f\left(1-\frac{y+1}{x}+\frac{3y(y+1)}{2x^2}+\order{\frac{1}{x^3}}\right)\,.\label{large_x_xy}
\end{align}
Using \eqref{appendix_entropy_dimless_variables} and \eqref{dimensionless_variables}, one finds
\begin{align}
    B=S_f-\frac{64\pi^2M_P^6}{T^2}+\frac{128\pi^2M_P^8(V_t+V_f)}{T^4}+\order{1}\,.\label{appendix_large_x_NLO}
\end{align}
We see that the dependence of the true vacuum energy density only enters at third order.

\section{Tunneling rates for KKLT and LVS type vacua}\label{appendix_sect_tunneling_calculations}
We recall that, if two vacua are connected by a barrier in a scalar-field potential $V(\phi)$, then the tension of the corresponding domain wall reads
\begin{align}
    T=\int_{\phi_f}^{\phi_t} d\phi\, \sqrt{2V(\phi)}\,. \label{Tension_def}
\end{align}
Here $\phi_f$ and $\phi_t$ are the field values characterizing the false and true vacuum. Parametrically, the tension obeys
\begin{align}
    T\sim \Delta \phi \sqrt{V_{barrier}}\,,\label{Tension_def_approx}
\end{align}
where $\Delta \phi$ is the width and $V_{barrier}$ the height of the potential barrier.

\paragraph{Decompactification:}
This tunneling transition generically exists due to the typical runaway behavior of the potential at field-space infinity. Equations \eqref{Tension_def}, \eqref{Tension_def_approx} still hold as an approximation, with $\phi_t$ being replaced by the field value at which the runaway potential has fallen to the value of the false vacuum. The width and the height of the barrier and hence the tension $T$ have been estimated in \cite{Kachru:2003aw,Westphal:2007xd}.
For KKLT, the barrier height is parametrically given by $\sim M_P^4 g_s|W_0|^2/\vol^2$, where $\vol$ is measured in 10d Planck units . The width is \cite{Westphal:2007xd}
\begin{align}
	\Delta\phi\sim \frac{M_P}{\vol^{2/3}}\,,
\end{align}
implying
\begin{align}
	T_{\text{dec, KKLT}}\sim \frac{M_P}{\vol^{2/3}}\frac{g_s^{1/2}M_P^2|W_0|}{\vol}=\frac{M_P^3g_s^{1/2}|W_0|}{\vol^{5/3}}\,.
\end{align}
Using the definition \eqref{dimensionless_variables} and assuming that $V_f-V_t=V_f\sim M_P^4 g_s|W_0|^2/\vol^2$, we find
\begin{align}
	x_{\text{dec, KKLT}}\sim \frac{M_P^6g_s|W_0|^2 \vol^2}{\vol^{10/3}M_P^6g_s|W_0|^2}\sim \frac{1}{\vol^{4/3}} < 1\label{x_dec_KKLT}\,.
\end{align}
According to \eqref{appendix_Minkowski_tunneling}, the tunneling exponent is given by
\begin{align}
    B=\frac{S_f}{(1+1/x_{\text{dec, KKLT}})^2}\,, \label{B_dec_KKLT}
\end{align}
and is thus \textbf{a few orders of magnitude} smaller than the entropy. Here we also recall that ${\cal V}\sim \ln(1/|W_0|)$ and hence, even if $W_0$ is tuned very small, can never become extremely large.
We also note that, if the uplifted cosmological constant is tuned to be very close to zero, such that $V_f$ is parametrically smaller than $g_s|W_0|^2/\vol^2$, the estimate \eqref{x_dec_KKLT} is no longer valid.
From the definition of $x$ \eqref{dimensionless_variables} (or directly from \eqref{appendix_Minkowski_tunneling}), one can see that a small $V_f$ in comparison to the barrier height leads to large $x$ and hence to the tunneling exponent being very close to the entropy.
However, the decay time cannot exceed the recurrence time \cite{Kachru:2003aw}.

For LVS models, the barrier height is $\sim M_P^4g_s^{1/2}|W_0|^2/\vol^3$ and the width is $\order{1}M_P$. As a result, the tension of the domain wall to decompactification scales as \cite{Westphal:2007xd}
\begin{align}
	T_{\text{dec, LVS}}\sim \frac{M_P^3g_s^{1/4}|W_0|}{\vol^{3/2}}\,.
\end{align}
Generically, the parameter $x$ and the decay exponent $B$ are then given by
\begin{align}
	x_{\text{dec, LVS}}\sim {\cal O}(1)\,,\qquad B=\frac{S_f}{(1+1/x_{\text{dec, LVS}})^2}\,.\label{B_dec_LVS}
\end{align}
We conclude that the decay exponent is \textbf{of the same magnitude} as the entropy. As before, tuning the vacuum energy to be very close to zero results in $x_\text{dec, LVS}\gg 1$ and hence in a maximally suppressed decay rate.

\paragraph{Flux transitions:}
The domain walls relevant for flux transitions in the type IIB landscape are given by D5/NS5 branes wrapping 3-cycles of the Calabi-Yau manifold. For LVS models, this has been discussed in \cite{deAlwis:2013gka}.
In units where $2\pi\sqrt{\alpha'}=1$, the D5 tension is $T_{\rm D5}=2\pi/g_s$. Using $M_P^2=4\pi\vol/g_s^{1/2}$ and wrapping the D5 on a $3$-cycle $\Sigma$, the domain wall tension reads\footnote{We use the convention that $M_P\int_{CY_3}\sqrt{g}=1=M_P\int_{CY}\Omega\wedge\overline{\Omega}$. Moreover, our volume ${\cal V}$ in 10d Planck units is related to the volume in string units by
$\tilde{\mathcal{V}}={\cal V}g_s^{3/2}$. 
}
\begin{align}
 T_{\text{D5, D.W.}}=M_P^3\frac{2\pi g_s^{1/2}}{(4\pi)^{3/2}\vol}\int_\Sigma \sqrt{g_\Sigma}\,.
\end{align}
If the 3-cycle $\Sigma$ were one of the special Lagrangian cycles $A^a$ or $B_b$ of a symplectic basis $(A^a,B_b)$, the last integral would be exactly equal to the absolute value of the corresponding period,
\begin{align}
	z^a=\int_{A^a}\Omega\,,\qquad \mathcal{G}_b=\int_{B_b}\Omega\,.
\end{align}
In general, $\Sigma$ is of the form
\begin{align}
	\Sigma = \sum_i(r^i_BB_i+s_i^AA^i)\,,\quad r^i_B,s_i^A\in\Z\,,
\end{align}
and the volume of $\Sigma$ can be given an upper and lower bound\footnote{Here we differ from \cite{deAlwis:2013gka}, where it is stated that the volume of the cycle $\Sigma$ is always given by the left hand side of \eqref{bound_volume_3cycle}. 
This difference does not parametrically affect the subsequent analysis of \cite{deAlwis:2013gka}. The reason is that they focus on the LVS, where $|W_0|\sim \order{{1}}$ before and after the transition and hence the lower and upper bound in \eqref{bound_volume_3cycle} are comparable. However, this difference is crucial for KKLT models, where both values of $|W_0|$ are tiny and hence the lower bound is much smaller than the upper bound.}
\begin{align}
	\left|\sum_i \left(r^i_B\mathcal{G}_i+s_i^Az^i\right)\right|\leq \int_\Sigma \sqrt{g_\Sigma}\leq \sum_i \left(|r^i_B\mathcal{G}_i|+|s_i^Az^i|\right)\,.\label{bound_volume_3cycle}
\end{align}
The D5 brane changes the $F_3$ flux on the cycles $A^a$ and $B_b$ precisely by $s_a^A$ and $r^b_B$ units such that the change in the flux superpotential becomes
\begin{align}
	\Delta W =\sum_i(r^i_B\mathcal{G}_i+s_i^Az^i)\,.
\end{align}
Hence, the domain wall tension satisfies the inequality
\begin{align}
	T_{\text{D5, D.W.}}=\frac{M_P^32\pi g_s^{1/2}}{(4\pi)^{3/2}\vol}\int_\Sigma \sqrt{g_\Sigma}\geq\frac{M_P^32\pi g_s^{1/2}}{(4\pi)^{3/2}\vol}\left|\Delta W\right|\,.\label{tension_bound_D5}
\end{align}

If, by analogy, the domain wall is built using an NS5 brane, the $H_3$ flux jumps and the superpotential changes by
\begin{align}
	\Delta W = \sum_i S\,(r^i_B\mathcal{G}_i+s_i^Az^i)\,,
\end{align}
with $S$ the axio-dilaton.
The corresponding relation for the tension is
\begin{align}
	T_{\text{NS5, D.W.}}=\frac{M_P^32\pi}{(4\pi)^{3/2}g_s^{1/2}\vol}\int_\Sigma \sqrt{g_\Sigma}\geq\frac{M_P^32\pi}{(4\pi)^{3/2}g_s^{1/2}\vol}\left|\Delta W/S\right|\,.\label{tension_bound_NS5}
\end{align}
We focus on the regime where $g_s$ is small such that $|S|\sim 1/g_s$.
As a result, for both the D5 and NS5 case, we have parametrically $T\gtrsim g_s^{1/2}M_P^3|\Delta W|/\vol$. However, this is not too important since generic domain walls are not near-BPS (see e.g.~\cite{Kachru:2002ns}) and hence do not saturate the lower bounds of 
\eqref{tension_bound_D5},\eqref{tension_bound_NS5}. Instead, a more useful estimate is provided simply by $T\gtrsim g_s^{1/2}M_P^3/\vol$.

Let us first focus on LVS models and their dS-dS transitions, with $|W_0|$ generically an ${\cal O}(1)$ number and ${\cal V}\sim \exp(1/g_s)$ exponentially large.
The requirement that $g_s$ remains small in the new vacuum constrains possible flux transitions.
The volume can change during the transition and for down-tunneling events it generally increases.
In this case, assuming also that $|W_0|$ does not change drastically, we estimate $V_f-V_t\sim V_f \sim M_P^4 g_s^{1/2}|W_0|^2/\vol^3$, with $\vol$ from now on always denoting the volume of the decaying vacuum. 
The parameter $y$ is then $\order{1}$ and $x$ may be estimated according to
\begin{align}
	x_{\text{LVS, flux}}\gtrsim \frac{g_s^{1/2}\vol}{|W_0|^2}\gg 1\,.\label{x_flux_LVS}
\end{align}
Using \eqref{large_x_xy},\eqref{appendix_large_x_NLO}, we obtain
\begin{align}
	B_{\text{LVS, flux}}&=S_f\left(1-\frac{y+1}{x_{\text{LVS, flux}}}+\order{\frac{1}{x_{\text{LVS, flux}}^2}}\right)\\
 &=S_f-\frac{64\pi^2M_P^6}{T^2}+\order{\vol}\sim \frac{24\pi^2\vol^3}{g_s^{1/2}|W_0|^2}-\frac{16(4\pi)^3\vol^2}{g_s\left(\int_\Sigma \sqrt{g_\Sigma}\right)^2}+\order{\vol}\,.\label{app_B_LVS_flux}
\end{align}
Here, $S_f$ denotes the entropy of the false vacuum.
At leading order,
\begin{align}
	B_{\text{LVS, flux}}\sim S_f\,.
\end{align}
We note that the dependence on the true vacuum energy density $V_t$ only enters at third order (see \eqref{appendix_large_x_NLO}).

For dS-dS transitions in KKLT models, both $|W_0|$ and $|W_0+\Delta W|$ must be tuned exponentially small. Thus, only a tiny fraction of flux transitions from a KKLT vacuum end in another KKLT dS state. This amounts to an extremely strong restriction on the allowed flux changes. As before, we have $T\sim M_P^3 g_s^{1/2}/\vol$.
For down-tunneling events, $|W_0|$ generically decreases during the transition. 
Using again the approximation $V_f-V_t\sim V_f \sim M_P^4 g_s|W_0|^2/\vol^2$, we find
\begin{align}
	x_{\text{KKLT, flux}}\sim \frac{1}{|W_0|^2}\gg 1\,,\label{x_flux_KKLT}
\end{align}
with $|W_0|$ the value of the flux superpotential of the parent vacuum.
Hence, by \eqref{large_x_xy},\eqref{appendix_large_x_NLO} with $y\sim 1$ due to our assumptions, 
\begin{align}
	B_{\text{KKLT, flux}} &=S_f\left(1-\frac{y+1}{x_{\text{KKLT, flux}}}+\order{\frac{1}{x_{\text{KKLT, flux}}^2}}\right)\\ 
    &=S_f-\frac{64\pi^2M_P^6}{T^2}\sim \frac{24\pi^2\vol^2}{g_s|W_0|^2}-\frac{16(4\pi)^3\vol^2}{g_s\left(\int_\Sigma \sqrt{g_\Sigma}\right)^2}\,.\label{app_B_KKLT_flux}
\end{align}
As for LVS models, the dependence on the true vacuum energy density $V_t$ only enters at third order.

\paragraph{KPV decay:} For completeness, we review the derivation of the KPV decay rate \cite{Kachru:2002gs,Freivogel:2008wm}. The $\overline{D3}$-brane of the KPV SUSY breaking mechanism can annihilate against the flux of the Klebanov-Strassler throat \cite{Kachru:2002gs}. This KPV or brane-flux annihilation process leads to the decay of dS models employing the $\overline{D3}$-brane uplift. The corresponding domain wall tension follows from the action of an NS5 brane wrapping the $S^3$ at the tip of the throat. In string frame with $2\pi\sqrt{\alpha'}=1$, it reads
\begin{align}
	T_{\rm KPV}=\frac{2\pi}{g_s^2}{\rm vol}(S^3)h_{\rm tip}^{-3/4}\tilde{\vol}^{1/2}=\frac{(bM)^{3/2}\tilde{\vol}^{1/2}}{2g_s^{1/2}}h_{\rm tip}^{-3/4}\,.
\end{align}
Here $b\approx 0.932$ is a numerical constant, $h_{\rm tip}^{-1/4}\sim \exp(-\frac{2\pi K}{3g_sM})$ is the warp factor at the tip of the throat, $M$ and $K$ are the flux numbers on the $A$ and $B$-cycle of the throat, and $\tilde{\vol}$ is the CY volume measured in string units.
The energy density provided by $N_{\overline{D3}}$ anti-branes at the tip of the throat is given by
\begin{align}
	\Delta V=\frac{4\pi N_{\overline{D3}}}{g_s}h_{\rm tip}^{-1}\tilde{\vol}^{2/3}\,.\label{energy_anti_D3}
\end{align}
Using $M_P^2=4\pi\tilde{\vol}/g_s^2$ and employing \eqref{dimensionless_variables} we find
\begin{align}
	x_{\rm KPV}=\frac{3b^3g_s^2M^3}{256 \pi^2\tilde{\vol}^{2/3} N_{\overline{D3}}}h_{\rm tip}^{-1/2}\,.
\end{align}
This formula is dominated by the exponentially small warp factor, ensuring that $\mbox{$x_{\rm KPV}\ll 1$}$. Since $y_{\rm KPV}\sim 1$, the KPV tunneling process occurs in the field-theoretic regime and represents the dominant decay channel. As a result, all dS vacua decay predominantly to AdS. This ensures that the assumptions which went into deriving the perturbative solution \eqref{p_i_solution} hold.
In the regime of $x_{KPV}\ll 1$, the KPV tunneling exponent follows from \eqref{decay_rate} and reads \cite{Kachru:2002gs,Freivogel:2008wm} 
\begin{align}
	B_{\rm KPV}= \frac{27b^6M^6g_s}{2048\pi (N_{\overline{D3}})^3}\,.\label{B_KPV}
\end{align}
Since 4d gravity plays no role, one expects the decay exponent not to depend on volume or warp factor but only on local physics at the tip of the throat. This is indeed the case.
\bibliographystyle{utphys}
\bibliography{References}
\end{document}